\documentstyle[12pt,aps,floats,epsf]{revtex}

\def\Journal#1#2#3#4{{#1} {\bf #2}, #3 (#4)}

\def\NPB{{\em Nucl. Phys.} B}
\def\PLB{{\em Phys. Lett.}  B}

\def\PRD{{\em Phys. Rev.} D}

\begin{document}

\begin{flushright}
JLAB-THY-00-05 \\
\end{flushright}

\vspace{0.5cm}

\centerline{\bf Deeply virtual Compton scattering at small x}

\vspace{1cm}
\centerline{IAN  BALITSKY and ELENA KUCHINA}
\vspace{0.5cm}
\centerline{ Physics Department, Old Dominion University, Norfolk 
VA 23529} 
\centerline{and}
\centerline{Theory Group, Jefferson Lab, Newport News VA 23606}
\centerline{e-mail: balitsky@jlab.org} 
\centerline{e-mail: kuchina@jlab.org}
\vspace{1cm}

\centerline{\bf Abstract} 

\vspace{0.4cm}

We calculate the cross section of the deeply virtual Compton scattering
at large energies and intermediate momentum transfers.

\bigskip
 PACS numbers: 12.38.Bx, 13.85.Fb, 13.85.Ni.
\newpage

\section{Introduction}
\bigskip

In recent years the study of the deeply virtual Compton scattering
(DVCS) became one of the most popular topics in QCD due to the fact that
it is determined by skewed parton distributions [1-3]
which generalize usual parton densities introduced by Feynman. 
These new probes of the nucleon structure are accessible in exclusive 
processes such as DVCS and potentially they can give us more information 
than the traditional parton densities. In this paper we consider the 
small-x DVCS 
where the energy of the incoming virtual photon $E$ is 
very large in comparison to its virtuality $Q^2$. 
(The first study
of the small-x DVCS was undertaken in Ref. \cite{bartels}). 
To be specific, we calculate
the DVCS amplitude in the region 
\begin{equation}
s\gg Q^2\gg -t\gg m^2
\label{fla1}
\end{equation}
 where $s=2mE$, $m$ is
the nucleon mass, and $t$ is the momentum transfer. The DVCS in this region is 
a semihard processes which can be described by the BFKL
(Balitsky-Fadin-Kuraev-Lipatov) pomeron \cite{bfkl}. It turns out that at large momentum transfer 
the coupling of the BFKL pomeron to the nucleon is essentially equal to the 
Dirac form factor of the nucleon $F_1(t)$, so the DVCS amplitude in the
region (\ref{fla1}) can be calculated without any model assumptions. The
results obtained in this region
can be used for the estimates of the
amplitude at experimentally accessible energies  where one or more conditions
in Eq. (\ref{fla1}) are relaxed. To be specific,  we have in mind the HERA
kinematics where $x\sim 10^{-2}\div 10^{-4}$,  $Q^2\geq 6$ GeV$^2$, and $-t\sim
1\div 5$ GeV$^2$ \cite{zeus}.  Since there
are only model predictions for the small-x DVCS in current literature
\cite{strikfurt1}, even the approximate calculations of the
cross section in QCD are very timely.

\section{Small-x DVCS in the lowest order in perturbation theory}
Similarly to the case of deep inelastic scattering (DIS),  the amplitude 
of DVCS is determined by the matrix element \cite{guvan}
\begin{equation}
H^{AB}=ie^{A}_{\nu}e^{B}_{\mu}
\int dz e^{iq'z}\langle p'| T\{j^{\mu}(z)j^{\nu}(0)\} |p\rangle
\label{fla2}
\end{equation}
where $q,p$ and $q',p'$ are the initial and the final momenta of the 
photon and the nucleon, respectively. The momentum transfer is defined as
$r=p'-p$. Since $Q^2=-q^2$ is large we can use perturbation theory for the
hard part of the DVCS process \cite{rad2}\cite{collins}. The typical diagram 
for the DVCS amplitude 
in the lowest order in perturbation theory is shown in Fig.1 
(recall that 
the diagrams with gluon exchanges dominate at high energies).
\begin{figure}[htb]
\hspace{4cm}
\mbox{
\epsfxsize=18cm
\epsfysize=9cm
\hspace{4cm}
\epsffile{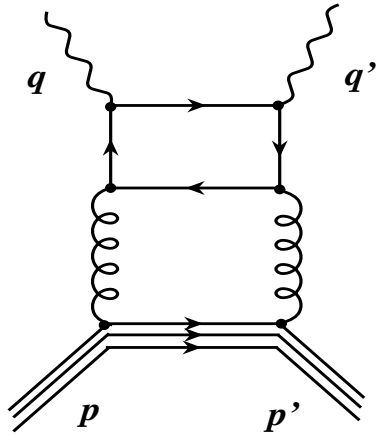}}
\vspace{-2.5cm}
{\caption{\label{fig1} A typical Feynman diagram for the high-energy 
$\gamma^*p\rightarrow\gamma p$ scattering}}
\end{figure}
It is convenient to calculate at first the imaginary part of the 
amplitude $H^{AB}$
\begin{equation}
V^{AB}={1\over \pi}{\rm Im} T^{AB}
\label{fla3}.
\end{equation}
In the leading order in perturbation theory 
 the amplitude at high energy is purely imaginary up to the  
 ${Q^2\over s}$ corrections (see e.g. the review \cite{lobzor}).
 At high orders in perturbation theory the amplitude will be 
purely imaginary in the leading logarithmic approximation (LLA)
 and we will
restore the real part using the dispersion relations.

\begin{figure}[htb]
\hspace{0cm}
\mbox{
\epsfxsize=12cm
\epsfysize=8cm
\hspace{1cm}
\epsffile{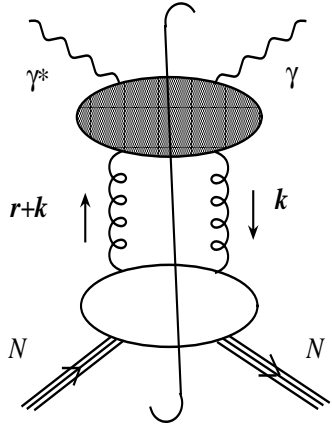}}
\vspace{-1cm}
{\caption{\label{fig2} Block structure of small-x DVCS in the 
leading order in perturbation theory}}
\end{figure}

\begin{figure}[htb]
\hspace{4cm}
\mbox{
\epsfxsize=16cm
\epsfysize=8cm
\hspace{0cm}
\epsffile{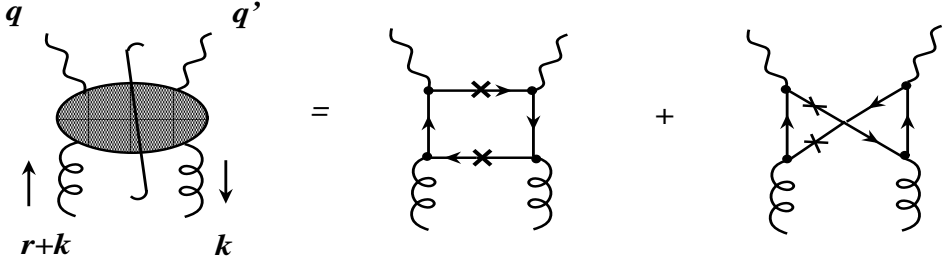}}
\vspace{-4cm}
{\caption{\label{fig3} Photon impact factor}}
\end{figure}

At high energies it is convenient to use the Sudakov variables. Let us 
define the light-like vectors $p_1= q'$,
$p_2=p'-{m^2\over s}p_1$, then 
\begin{eqnarray}
q&=&p_1(1-{r_{\perp}^2\over s})-x p_2-r_{\perp}~~~~~~~~~~~~~~q'=p_1
\nonumber\\
p&=&p_2(1+x)+{m^2+r_{\perp}^2\over s}p_1+r_{\perp}~~~~~~~~~~~~~~~~
p'=p_2+{m^2\over s} p_1
\label{fla4}
\end{eqnarray}
where $x\equiv{Q^2+t\over s}\simeq {Q^2\over s}=x_{Bj}$
and $t\simeq -r_{\perp}^2$ at large energies. 
Consider the integral over gluon momentum $k=\alpha_kp_1+
\beta_kp_2+k_{\perp}$
\begin{equation}
V^{AB}={2\over\pi}g^4\int {d^4k\over 16\pi^4} {1\over k^2}
{1\over (r+k)^2}{\rm Im} \Phi^{ab}_{\xi\eta}(k+r, -k) 
{\rm Im}\Phi^{\xi\eta ab}_N(-k-r,k)
\label{fla5}
\end{equation}
where $\Phi^{ab}_{\xi\eta}(k, r+k)$ and $(\Phi_N)^{ab}_{\xi\eta}(k,r+k)$ are 
the upper and the lower blocks 
of the diagram in Fig. \ref{fig2} (stripped of the strong 
coupling constant $g$). 
Here $a,b$ and $\xi,\eta$ are the color and
Lorentz indices, respectively.
It is well known that in the Regge kinematics ($\equiv~s\gg$ everything else) 
$\alpha_k\sim {m^2\over s}$, 
and $\beta_k\sim x$ so $k^2\simeq -k_{\perp}^2$. Moreover, 
$\alpha$'s in the upper block are $\sim 1$ 
so one can drop $\alpha_k$ in the upper block. Similarly, $\beta$'s in the 
lower block are $\sim 1$ 
and one can neglect $\beta_k$ in the lower block. We get 
($\Phi^{ab}={\delta_{ab}\over 8}\Phi^{cc}$): 
\begin{equation}
V^{AB}={g^4\over 4\pi}\int {d^4k\over 16\pi^4} {1\over k_{\perp}^2}
{1\over (r+k)_{\perp}^2}{\rm Im} 
\left.\Phi^{aa}_{\xi\eta}(k+r, -k)\right|_{\alpha_k=0} 
{\rm Im}\left.\Phi^{\xi\eta bb}_N(-k-r,k)\right|_{\beta_k=0}
\label{fla6}.
\end{equation}
At high energies, the metric tensor in the numerator of the  Feynman-gauge 
gluon propagator reduces to 
$g^{\mu\nu}\rightarrow {2\over s}p_2^{\mu}p_1^{\nu}$
so the integral (\ref{fla6}) for 
the imaginary part factorizes into a product of two ``impact factors"
integrated with two-dimensional propagators
\begin{equation}
V^{AB}={2s\over \pi}g^4\left(\sum e_q^2\right)
\int {d^2k_{\perp}\over 4\pi^2} {1\over k_{\perp}^2}
{1\over (r+k)_{\perp}^2}I(k_{\perp},r_{\perp}) I_N(k_{\perp},r_{\perp})
\label{fla7}
\end{equation}
where
\begin{eqnarray}
I(k_{\perp},r_{\perp})&=&{1\over 2s}p_2^{\xi}p_2^{\eta}
\left(\sum e_q^2\right)
\left.\int {d\beta_k\over 2\pi}
{\rm Im} \Phi^{aa}_{\xi\eta}(k+r, -k)\right|_{\alpha_k=0}
\label{fla8}\\
I_N(k_{\perp},r_{\perp})&=&{1\over 2s}p_1^{\xi}p_1^{\eta}
\left.\int {d\alpha_k\over 2\pi}
{\rm Im} \Phi^{aa}_{N\xi\eta}(-k-r,k)\right|_{\beta_k=0}
\label{fla9}
\end{eqnarray}
and $\left(\sum e_q^2\right)$ is the sum of squared charges of active flavors 
($u,d,s$, and possibly $c$). 
The photon impact factor is given by the two one-loop diagrams shown in 
Fig. \ref{fig3}.
The standard calculation of these diagrams\cite{ifak} yields
\begin{equation}
{I}^{AB}(k_{\perp},r_{\perp})=\bar{I}^{AB}(k_{\perp},r_{\perp})- 
\bar{I}^{AB}(0,r_{\perp})
\label{fla10}
\end{equation}
where
 \begin{eqnarray}   
&&\bar{I}^{AB}(k_{\perp},r_{\perp}) =
   {1\over 2}\int _{0}^{1} \frac {d\alpha}{2\pi }
   \int _{0}^{1} \frac {d\alpha'}{2\pi }
   \left\{
   P_{\perp}^{2}\alpha'\bar{\alpha'}+Q^2\alpha'\alpha\bar{\alpha}
   \right\}^{-1}
\label{fla11}\\
&& 
   \Bigg\{
      (1-2\alpha\bar{\alpha})
      P_{\perp}^{2} (e^A, e^B)_{\perp}
+4\alpha\bar{\alpha}\bar{\alpha'}[P_{\perp}^{2} (e^A, e^B)-
2(e^A, P)_{\perp}(e^B, P)_{\perp}]
-4\alpha\bar{\alpha}(1-2\alpha)(r,e^A)_{\perp} (P,e^B)_{\perp}   
\Bigg\} \nonumber
\end{eqnarray}
for the transverse polarizations $A,B=1,2$ (cf. \cite{ing}) and
\begin{eqnarray} 
&&  \bar{I}^{3B}(k_{\perp},r_{\perp}) =
   {1\over 2Q}\int _{0}^{1} \frac {d\alpha}{2\pi }
   \int _{0}^{1} \frac {d\alpha'}{2\pi }
   \left\{
   P_{\perp}^{2}\alpha'\bar{\alpha'}+Q^2\alpha'\alpha\bar{\alpha}
   \right\}^{-1}
\label{fla12}\\
&& 
   \Bigg\{
      (1-2\alpha\bar{\alpha})
      P_{\perp}^{2} (r, e^B)_{\perp}
+4\alpha\bar{\alpha}\bar{\alpha'}[P_{\perp}^{2} (r, e^B)_{\perp}-
2(r, P)_{\perp}(e^B, P)_{\perp}]
      -4\alpha\bar{\alpha}(1-2\alpha)Q^2(P,e^B)_{\perp}
   \Bigg\} \nonumber
\end{eqnarray}
 for the longitudinal polarization 
 \begin{equation}
 e^3(q)={1\over Q}(p_1+xp_2)
 \label{fla13}
 \end{equation}
. Here $P_{\perp}\equiv k_{\perp}+r_{\perp}\alpha$ and $(a,b)_{\perp}$ 
denotes the (positive) scalar
product of transverse components of vectors $a$ and $b$. At
large transverse momenta $k_{\perp}^2\gg r_{\perp}^2$ the impact factor 
(\ref{fla10})
reduces to
\begin{equation}
 I^{AB}(k_{\perp},r_{\perp})\rightarrow 
 {(e^A, e^B)_{\perp}\over 4\pi^2}{k_{\perp}^2\over Q^2}
 \ln{Q^2\over r_{\perp}^2}
\label{fla14}.
\end{equation}

The impact factor for the proton which decribes the pomeron-nucleon
coupling cannot be calculated in the 
perturbation theory. However, in the next section we demonstrate that at high 
momenta $k_{\perp}^2\gg m^2$ this impact factor reduces to
\begin{equation}
I_N(k_{\perp},r_{\perp})\stackrel{k_{\perp}^2\gg m^2}{=}
F_1^{p+n}(t)
\label{fla15}
\end{equation}
where $F_1^{p+n}(t)$ is the sum of the proton and neutron 
Dirac form factors. As we shall see below, 
the characteristic transverse momenta in our 
gluon loop are large so the estimate (\ref{fla15}) is sufficient for our 
purposes. Substituting the
nucleon impact factor (\ref{fla15}) into Eq. (\ref{fla7}) we obtain
\begin{equation}
V^{AB}={2s\over \pi}g^4(\sum e_q^2)
F_1^{p+n}(t)
\int {d^2k_{\perp}\over 4\pi^2} 
{I^{AB}(k_{\perp},r_{\perp})\over k_{\perp}^2(r+k)_{\perp}^2}
\label{fla16}.
\end{equation}
Performing the final integration over $k_{\perp}$, one gets
\begin{eqnarray}
\lefteqn{ V^{AB}=}\nonumber\\
&&{2\over x}\left({\alpha_s\over \pi}\right)^2(\sum_{\rm flavors}e_q^2) 
F_1^{p+n}(t)\nonumber\\
&&
\Bigg ( (e^A,e^B)_{\perp}\left({1\over 2} \ln^2{Q^2\over |t|}+2\right)-
(e^A,e^B)_{\perp}+{2\over r_{\perp}^2}(e^A,r)_{\perp}(e^B,r)_{\perp} 
+O(t/Q^2)\Bigg )\label{fla17}
\end{eqnarray}
for the transverse polarizations and
\begin{eqnarray}
\lefteqn{ V^{3B}=}\nonumber\\
&&-{2\over x}\left({\alpha_s\over \pi}\right)^2(\sum_{\rm flavors}e_q^2) 
F_1^{p+n}(t)
{(r,e^B)_{\perp}\over Q}
\left({1\over 2} \ln^2{Q^2\over |t|}-5\ln{Q^2\over |t|}+
{15\over 2}-{\pi^2\over 3}+O(t/Q^2)\right)
\label{fla18}
\end{eqnarray}
for the longitudinal one. The longitudinal amplitude (\ref{fla18}) 
is twist-suppressed as ${\sqrt{|t|}\over Q}$ in comparison to 
the transverse amplitude (\ref{fla17}) (as it should, due to the fact that
$t\rightarrow 0$ corresponds to real incoming photon). 

Since the integral
over $k_{\perp}$ (\ref{fla16}) converges at $k_{\perp}\sim Q$ the region 
$k_{\perp}\sim m$,
where we do not know the nucleon impact factor, contributes to the terms $\sim
O(t/Q^2)$ which we neglect.  

\section{Nucleon impact factor}

\begin{figure}[htb]
\vspace{-1cm}
\mbox{
\epsfxsize=16cm
\epsfysize=7cm
\hspace{0cm}
\epsffile{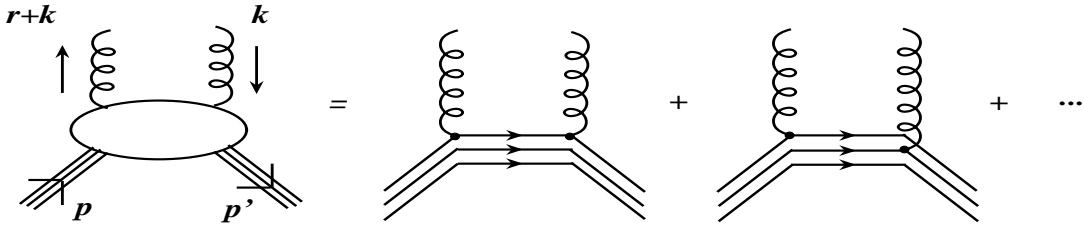}}
\vspace{-2cm}
{\caption{\label{fig4} Nucleon impact factor}}
\end{figure}

In the lowest order in perturbation theory there is no difference 
between the diagrams for the nucleon impact factor 
shown in Fig. \ref{fig4} 
and similar diagrams with two gluons 
replaced by two photons (up to  the trivial numerical factor 
$c_F={4\over 3}$ and 
replacement of $e\leftrightarrow g$). In this case the lower part of
the diagram can be formally written as follows:
\begin{equation}
\Phi_N(-k-r, k)\stackrel{\rm def}{\equiv} 
{1\over 2}{p_1^\xi p_1^\eta\over s}
(\Phi_N)_{\xi\eta}^{bb}(-k-r, k)=
{2\over 3}i p_1^{\mu} p_1^{\nu}\int dz e^{ikz}
\langle p'|T^*\{J_{\mu}(z)J_{\nu}(0)\}| p\rangle
\label{fla19}
\end{equation}
where $J_{\mu}=\bar{u}\gamma_{\mu}u+\bar{d}\gamma_{\mu}d $. 
The  $T^*$ means the T-product where the diagrams with 
pure gluon exchanges in t-channel are excluded; by definition, such diagrams 
contribute to subsequent ranks of BFKL ladder rather than to impact 
factor. (This is the reason why we have not included in $J$ the contribution 
of strange quarks). 
Since $k^2$ in 
our case is large and negative (-$k^2=k_{\perp}^2\gg m^2$) we can expand
the T-product of two currents near the light cone (see e.g. \cite{bal83}) 
\begin{equation}
\Phi_N(k, r+k)=
{2\over 3s}\int dz e^{ikz}{zp_1\over \pi^2 z^4}
\langle p'|-\bar{\psi}(z)[z,0]\not\! p_1\psi(0)+\bar{\psi}(0)[0,z]\not\! p_1\psi(z)| p\rangle^*_{z^2=0}
\label{fla20}
\end{equation}
where again $\langle...\rangle^*$ stands for the matrix element with pure gluon 
exchanges excluded. Here $[x,y]$ denotes
the gauge link connecting the points $x$ and $y$ 
($[x,y]\equiv Pexp\left( ig\int_0^1 du(x-y)^{\mu}A_{\mu}(ux+(1-u)y\right)$). 
The matrix element (\ref{fla16}) can be parametrized
in terms of skewed parton distributions\cite{rad2} as follows
\begin{eqnarray}
&&\langle p',\lambda'|\bar{q}(z)[z,0]\not\! p_1q(0)| p,\lambda\rangle^*_{z^2=0}
=\label{fla21}\\
&&
\bar{u}(p',\lambda')\not\! p_1u(p,\lambda) 
\int_0^1 dX e^{i(X-x)pz}{\cal V}^q_x(X,t)+
{1\over 2m}\bar{u}(p',\lambda')\not\! p_1\!\not\! r_{\perp} u(p,\lambda)
\int_0^1 dX e^{i(X-x)pz}{\cal W}^q_x(X,t)\nonumber\\ 
&&
\langle p',\lambda'|\bar{q}(0)[0,z]\not\! p_1q(z)| p,
\lambda\rangle^*_{z^2=0}=\nonumber\\
&&
\bar{u}(p',\lambda')\not\! p_1u(p,\lambda) 
\int_0^1 dX e^{-iXpz}{\cal V}^q_x(X,t)+
{1\over 2m}\bar{u}(p',\lambda')\not\! p_1\!\not\! r_{\perp} u(p,\lambda)
\int_0^1 dX e^{-iXpz}{\cal W}^q_x(X,t)
\nonumber,
\end{eqnarray}
 where ${\cal V}^u_x(X,t)$ and ${\cal W}^u_x(X,t)$ are the nonflip and 
spin-flip 
 skewed parton distributions for 
the {\it valence} $u$ quark 
(recall that we must take into account only valence quarks since we forbid 
diagrams with pure gluon exchanges). Similarly, 
${\cal V}^d_x(X,t)$ and ${\cal W}^d_x(X,t)$ refer to the valence $d$-quark 
distributions. 
At large energies 
$\bar{u}(p',\lambda')\not\! p_1u(p,\lambda)=s\delta_{\lambda\lambda'}$,  so
\begin{eqnarray}
\lefteqn{\langle p',\lambda'|\bar{q}(0)[0,z]\not\! p_1q(z)-
\bar{q}(z)[z,0]\not\! p_1q(0)| p,\lambda\rangle^*_{z^2=0}=}\label{fla22}\\
&&
\int_0^1 dX \left(e^{-iXpz}- 
e^{i(X-x)pz}\right)\left[s\delta_{\lambda\lambda'}{\cal V}^q_x(X,t)+ 
{1\over 2m}\bar{u}(p',\lambda')\not\! p_1\!\not\! r_{\perp} u(p,\lambda)
{\cal W}^q_x(X,t)\right]
\nonumber.
\end{eqnarray}
After integration over $z$ the lower block (\ref{fla19}) reduces to
\begin{eqnarray}
\lefteqn{\Phi_N(-k-r, k)=}\label{fla23}\\
&&
{2\over 3s}
\int_0^1 dX 
\left[{(X-x)s+2p_1\cdot k\over -k^2-2p\cdot k (X-x)-i\epsilon}
-{-Xs+2p_1\cdot k\over -k^2+2p\cdot k X-i\epsilon}\right]
\nonumber\\
&&
\left(\delta_{\lambda\lambda'}({\cal V}^u_{x}(X,t)+{\cal V}^d_x(X,t))+
{1\over 2ms}\bar{u}(p',\lambda')\not\! p_1\!\not\! r_{\perp} u(p,\lambda)
({\cal W}^u_{x}(X,t)+{\cal W}^d_x(X,t)) \right)
\nonumber.
\end{eqnarray}
The nucleon impact factor (\ref{fla9}) is the integral of 
the imaginary part of
r.h.s. of eq. (\ref{fla23}) over energy
\begin{eqnarray}
\lefteqn{I_N(k_{\perp},r_{\perp})=\int_0^1 {d\alpha_k\over 2\pi}
{\rm Im}\Phi_N(-(\alpha_k-{r_{\perp}^2\over s}) p_1-k_{\perp}-
r_{\perp},\alpha_k p_1+k_{\perp} )=}\label{fla24}\\
&&{1\over 3}\int_0^1 d\alpha_k\int_x^1 dX \left[s(X-x)
\delta (k_{\perp}^2-\alpha_k s(X-x))-sX\delta (k_{\perp}^2+ \alpha_k sX)\right]
\nonumber\\
&&
\left(\delta_{\lambda\lambda'}({\cal V}^u_x(X,t)+{\cal V}^d_x(X,t))+
{1\over 2ms}\bar{u}(p',\lambda')\not\! p_1\!\not\! r_{\perp} u(p,\lambda)
({\cal W}^u_{x}(X,t)+{\cal W}^d_x(X,t))\right)
\nonumber\\
&&={1\over 3}\int_x^1 dX\left(\delta_{\lambda\lambda'}({\cal V}^u_x(X,t)+
{\cal V}^d_x(X,t))+{1\over 2ms}\bar{u}(p',\lambda')\not\! p_1\!\not\! r_{\perp} u(p,\lambda)
({\cal W}^u_{x}(X,t)+{\cal W}^d_x(X,t))\right)
\nonumber.
\end{eqnarray}
Since valence quark distributions decrease at $x\rightarrow 0$ 
we can extend the lower limit
of integration in r.h.s. of eq. (\ref{fla24}) to 0 and obtain
\begin{eqnarray}
\lefteqn{I_N(k_{\perp}, r_{\perp})\stackrel{k_{\perp}^2\gg m^2}{=}}
\label{fla25}\\
&&{1\over 3}
\int_0^1 dX
\left(\delta_{\lambda\lambda'}({\cal V}^u_x(X,t)+{\cal V}^d_x(X,t))+
{1\over 2ms}\bar{u}(p',\lambda')\not\! p_1\!\not\! r_{\perp} u(p,\lambda)
({\cal W}^u_{x}(X,t)+{\cal W}^d_x(X,t))
\right)
\nonumber.
\end{eqnarray}
Let us recall the sum rules \cite{ji},\cite{rad2}
\begin{eqnarray}
\int_0^1 dX
\left({\cal F}^q_x(X,t)-{\cal F}^{\bar q}_x(X,t)\right)&=&F^q_1(t)
\nonumber\\
\int_0^1 dX
\left({\cal K}^q_x(X,t)-{\cal K}^{\bar q}_x(X,t)\right)&=&F^q_2(t)
\label{fla26}
\end{eqnarray}
where ${\cal F}^q_x(X,t)$ and ${\cal K}^q_x(X,t)$ are the total 
(valence $+$ sea) 
nonflip and spin-flip skewed quark 
distributions while  ${\cal F}^{\bar q}_x(X,t)$ and 
${\cal K}^{\bar q}_x(X,t)$ are 
the antiquark ones. Here $F^q_1(t)$ and $F^q_2(t)$ stand for 
the $q$-quark 
components of the Dirac and Pauli
form factors of the proton). 
Since the contribution of sea quarks drops from the difference  
${\cal F}^q-{\cal F}^{\bar q}$ we can rewrite eqs. (\ref{fla26}) as the sum
rules for valence quark distributions
\begin{equation}
\int_0^1 dX {\cal V}^q_x(X,t)= F^q_1(t),~~~~~~~~~
\int_0^1 dX {\cal W}^q_x(X,t)= F^q_2(t)
\label{fla27}.
\end{equation}
Substituting this estimate to eq. (\ref{fla25}) and 
using the isospin invariance, we get the final result
for the nucleon impact factor at large transverse momenta
\begin{equation}
I_N(k_{\perp}, r_{\perp})\stackrel{k_{\perp}^2\gg m^2}{=}
\delta_{\lambda\lambda'}F_1^{p+n}(t)+
{1\over 2ms}\bar{u}(p',\lambda')\not\! p_1\!\not\! r_{\perp} 
u(p,\lambda)F_2^{p+n}(t)
\label{fla28},
\end{equation}
where 
$F_1^{p+n}(t)\equiv F_1^p(t)+F_1^n(t)$ and 
$F_2^{p+n}(t)\equiv F_2^p(t)+F_2^n(t)$.
As usual, $F_1^{p(n)}$ and 
$F_2^{p(n)}$ are the Dirac and Pauli form factors of 
the proton (neutron), respectively. 
With our accuracy they can be approximated
by the dipole formulas
\begin{equation}
\begin{array}{lllllll}
F_1^p+{t\over 4m^2}F_2^p&=&G_E^p=&{1\over \left(1+{|t|\over
0.7{\rm GeV}^2}\right)^2}~~~~~~~~~~~~~~& 
F_1^p+F_2^p=G_M^p&=&{2.79\over \left(1+{|t|\over 0.71{\rm GeV}^2}\right)^2}
\\
F_1^n+{t\over 4m^2}F_2^n&=&G_E^n=&0
&F_1^n+F_2^n=G_M^n&=&{-1.91\over \left(1+{|t|\over 0.71{\rm GeV}^2}\right)^2}
\end{array}
\label{fla29},
\end{equation}
which leads to
\footnote
{Literally, one obtains  
\begin{equation}
F_1^{p+n}(t)= {1\over 1+\left({|t|\over 0.71GeV^2}\right)^2}
{1+0.88{|t|\over 4m^2}\over 
1+{|t|\over 4m^2}},~~~~~~~~F_2^{p+n}= {0.12\over 1+
\left({|t|\over 0.71GeV^2}\right)^2}
\label{fla30}
\end{equation}
but with our accuracy we can use the estimate (\ref{fla31}).
} 
\begin{equation}
F_1^{p+n}(t)= {1\over 1+
\left({|t|\over 0.7GeV^2}\right)^2},~~~~~~~~F_2^{p+n}=0 
\label{fla31}
\end{equation}
. Note that the spin-flip term turned out to be negligible 
for our values of $t$.
Moreover, it vanishes at $t=0$ which suggests that it 
is numerically small at all $t$.

 Our final estimate of the nucleon 
impact factor is 
\begin{equation}
I_N(k_{\perp}, r_{\perp})\stackrel{k_{\perp}^2\gg m^2}{=}
\delta_{\lambda\lambda'}F_1^{p+n}(t)
\label{fla32}
\end{equation}
where $F_1^{p+n}$ is given by the dipole formula (\ref{fla31})
\footnote{
The dipole formula for the neutron form factor does not seem to work
as well as the dipole formula for the proton form factor. As a measure
of the uncertainty we can compare the results obtained from eq. (\ref{fla31})
to those obtained using the model from Ref. \cite{prof} (which
was fit only to the proton form factor)
\begin{eqnarray}
F^{p+n}_1(t)&=&
{1\over 3}\int_0^1dX\left({\cal V}^u_x(X,t)+{\cal V}^d_x(X,t)\right)\nonumber\\
{\cal V}^u_x(X,t)&=&1.89 X^{-0.4}\bar{X}^{3.5}(1+6X)
\exp\left(-{\bar{X}\over X}{|t|\over 2.8{\rm GeV}^2}\right)
\nonumber\\
{\cal V}^d_x(X,t)&=&0.54 X^{-0.6}\bar{X}^{4.2}(1+8X)
\exp\left(-{\bar{X}\over X}{|t|\over 2.8{\rm GeV}^2}\right)
\label{fla33}
\end{eqnarray}
The results for the DVCS cross section in this model are about 
$1.5$ times bigger than the results obtained from 
the dipole formula (\ref{fla31}).
}.
In what follows we shall omit the factor $\delta_{\lambda\lambda'}$ 
(as it was done
in eq. (\ref{fla15})) since all our
amplitudes will always be diagonal in the proton's spin. 
 
\section{The BFKL ladder}

\begin{figure}[htb]
\mbox{
\epsfxsize=18cm
\epsfysize=10cm
\hspace{3cm}
\epsffile{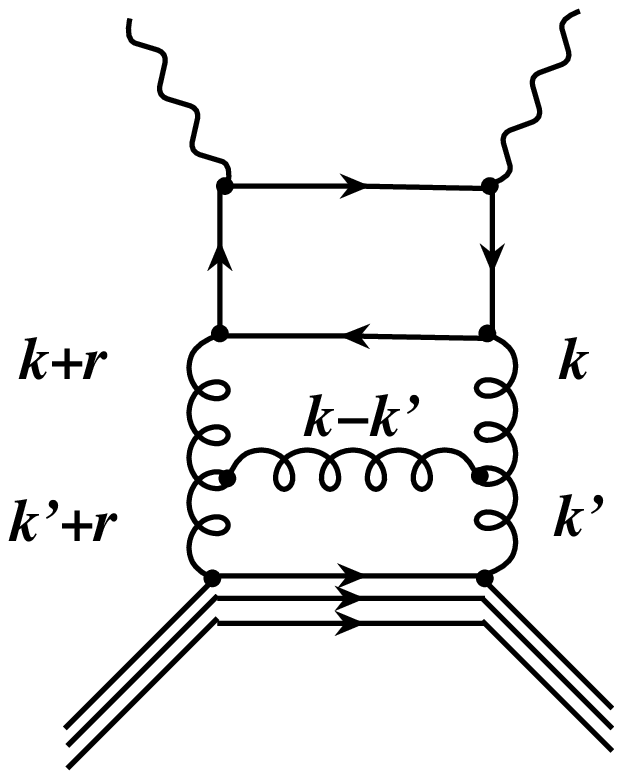}}
\vspace{-3cm}
{\caption{\label{fig5} Typical diagram in the next-to-leading order 
in perturbation theory}}
\end{figure}

In the next order in  perturbation theory the most important diagrams are 
those of the type shown in Fig. (\ref{fig5})
\footnote
{Actually, this diagram gives the total contribution in LLA if one 
replaces the three-gluon vertex in Fig. (\ref{fig5}) by the 
effective Lipatov's
vertex \cite{lobzor}} . 
Calculation of this diagrams in the leading log approximation yields
\begin{eqnarray}
 \lefteqn{V^{AB}=}
 \nonumber\\
 &&{2sg^4\over\pi}(\sum e_q^2)\left(6\alpha_s\ln {1\over x}\right)
 \int {d^2k_{\perp}\over 4\pi^2}{d^2k'_{\perp}\over 4\pi^2} 
  {I^{AB}(k_{\perp},r_{\perp})\over k_{\perp}^2(r+k)_{\perp}^2}
  K(k_{\perp},k'_{\perp},r_{\perp})
{I_N(k'_{\perp},r'_{\perp})\over (k'_{\perp})^{2} (r+k')_{\perp}^2}
\label{fla34}
\end{eqnarray}
where $K(k_{\perp},k'_{\perp},r_{\perp})$ is the BFKL kernel\cite{bfkl}
\begin{eqnarray}
\lefteqn{ K(k_{\perp},k'_{\perp},r_{\perp})=}\nonumber\\
& -r_{\perp}^2+{k_{\perp}^2(r-k')_p^2\over (k-k')_{\perp}^2}
+ {k_{\perp}^2(r-k')_p^2\over (k-k')_{\perp}^2} +
k_{\perp}^2(k-p)_{\perp}^2{1\over 2}\delta(k_{\perp}-k'_{\perp})\int dp_{\perp} 
\left({k_{\perp}^2\over p_{\perp}^2 (k-p)_{\perp}^2}+
{(k-r)_{\perp}^2\over(p-r)_{\perp}^2 (k-p)_{\perp}^2}\right)
\label{fla35}
\end{eqnarray}
As we shall see below, the integral over $k'_{\perp}$ converges at 
$|k'_{\perp}|\gg m$ so we can again use the approximation (\ref{fla15})
for the nucleon impact factor. One obtains
\begin{equation}
 \int d^2k'_{\perp}  
 K(k_{\perp},k'_{\perp},r_{\perp})
{1\over (k'_{\perp})^{2} (r+k')_{\perp}^2}I_N(k'_{\perp},r'_{\perp})=
\pi F_1^{p+n}(t)
\left(\ln {k_{\perp}^2\over r_{\perp}^2}+
\ln {(k-r)_{\perp}^2\over r_{\perp}^2}\right)
\label{fla36}
\end{equation}
and therefore the amplitude (\ref{fla34}) takes the form
\begin{equation}
 V^{AB}={g^4 s\over \pi}F_1^{p+n}(t)
 \left({3\alpha_s\over\pi}\ln {1\over x}\right)
 \int {d^2k_{\perp}\over 4\pi^2} 
 {I(k_{\perp},r_{\perp}) \over k_{\perp}^2(r+k)_{\perp}^2}
 \left(\ln {k_{\perp}^2\over r_{\perp}^2}+
\ln {(k-r)_{\perp}^2\over r_{\perp}^2}\right)
\label{fla37}.
\end{equation}
Finally, the integration over $k$ yields
\begin{eqnarray}
\lefteqn{ V^{AB}=}\nonumber\\
&&{2\over x}\left({\alpha_s\over\pi}\right)^2(\sum_{\rm flavors}e_q^2) 
F_1^{p+n}(t)\left({3\alpha_s\over\pi}\ln {1\over x}\right)\nonumber\\
&&
\Bigg ( (e^A,e^B)_{\perp}\left({1\over 6} \ln^3{Q^2\over |t|}+
2\ln{Q^2\over |t|}-2+\zeta(3)\right)+
\left({2\over r_{\perp}^2}(e^A,r)_{\perp}(e^B,r)_{\perp} -
(e^A,e^B)_{\perp}\right)
\Bigg )
\label{fla38}
\end{eqnarray}
where the accuracy is $O({1\over \ln x})$.

\begin{figure}[htb]
\hspace{4cm}
\mbox{
\epsfxsize=19cm
\epsfysize=12cm
\hspace{3cm}
\epsffile{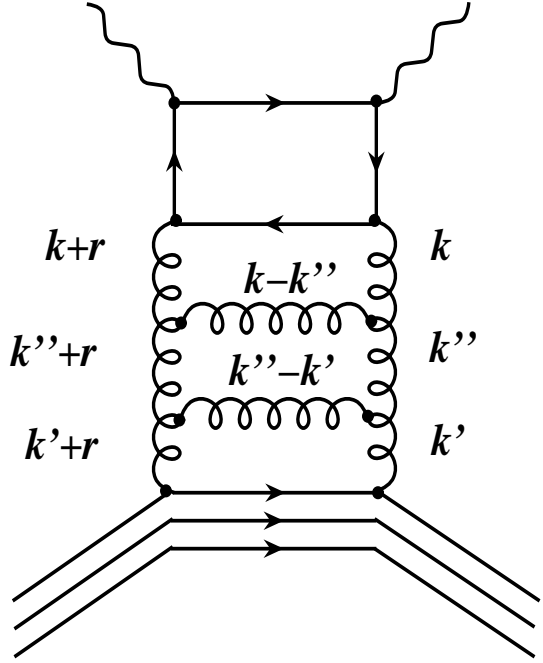}}
\vspace{-1cm}
{\caption{\label{fig6} Typical diagram in the next-to-next-to-leading order 
in perturbation theory}}
\end{figure}

In the next order in BFKL approximation (see Fig. \ref{fig6}) it is 
still possible  to obtain the DVCS amplitude (\ref{fla3}) in 
the explicit form (we have not obtained the explicit expressions for 
 higher-order terms in the BFKL expansion (\ref{fla38})
\footnote{It is possible to write down the result of the summation of the BFKL
ladder in the form of Mellin integral over complex momenta using the Lipatov's
conformal eigenfunctions of the BFKL equation in the coordinate space.
Unfortunately, we were not able to perform explicitly the integration of the
Lipatov's eigenfunctions with impact factors and without it the Mellin
representation  of the DVCS amplitude is useless for practical 
applications.}). 
The amplitude in this order is 
\begin{eqnarray}
 V^{AB}&=&{g^4s\over\pi}(\sum e_q^2)
\left(6\alpha_s\ln {1\over x}\right)^2
\int {d^2k_{\perp}\over 4\pi^2}{d^2k'_{\perp}\over 4\pi^2}
{d^2k"_{\perp}\over 4\pi^2}I(k_{\perp},r_{\perp})\label{fla39}\\
&&  {1\over k_{\perp}^2(r+k)_{\perp}^2}K(k_{\perp},k"_{\perp},r_{\perp}) 
{1\over (k"_{\perp})^{2}
(r+k")_{\perp}^2}K(k"_{\perp},k'_{\perp},r_{\perp})
{1\over (k'_{\perp})^2(r+k')_{\perp}^2}
I_N(k'_{\perp},r'_{\perp}) 
\nonumber.
\end{eqnarray}
Once again, if we use the fact that the integral over $k'_{\perp}$ 
converges at 
$|k'_{\perp}|\gg m$  we can approximate the nucleon impact factor
by eq. (\ref{fla32}), and obtain
\begin{eqnarray}
&&\int {d^2k'_{\perp} \over 4\pi^2} \int {d^2k"_{\perp} \over 4\pi^2} 
 K(k_{\perp},k"_{\perp},r_{\perp})
{1\over (k")_{\perp}^{2} (r+k")_{\perp}^2}K(k"_{\perp},k'_{\perp},r_{\perp})
{1\over (k')_{\perp}^{2} (r+k')_{\perp}^2}I_N(k'_{\perp},r'_{\perp})=
\nonumber\\
&&
{1\over 4\pi} F_1^{p+n}(t)
\int {d^2k"_{\perp} \over 4\pi^2} 
{K(k_{\perp},k"_{\perp},r_{\perp})\over (k")_{\perp}^{2} (r+k")_{\perp}^2}
\left(\ln {(k"_{\perp})^2\over r_{\perp}^2}+
\ln {(k"-r)_{\perp}^2\over r_{\perp}^2}\right)=\nonumber\\
&&
{1\over 16\pi^2}F_1^{p+n}(t)
\left(\ln^2 {k_{\perp}^2\over r_{\perp}^2}+
\ln^2 {(k-r)_{\perp}^2\over r_{\perp}^2}\right)
\label{fla40}.
\end{eqnarray}
The resulting integration over $k_{\perp}$ yie lds
\begin{eqnarray}
\lefteqn{ V^{AB}=}\nonumber\\
&&{9\over x}\left({\alpha_s\over\pi}\right)^4(\sum e_q^2) 
F_1^{p+n}(t)\ln^2x
\Bigg[  (e^A,e^B)_{\perp}\Big({1\over 24} \ln^4{Q^2\over |t|}+
\ln^2{Q^2\over |t|}
-2\ln{Q^2\over |t|}+\nonumber\\
&&2(\zeta(3)-1)+1.46 \Big)+
\left({2\over r_{\perp}^2}(e^A,r)_{\perp}(e^B,r)_{\perp} -
(e^A,e^B)_{\perp}\right)
\Bigg]
\label{fla41}.
\end{eqnarray}
As we mentioned, we were not able to obtain the explicit expressions
for the amplitude in higher orders in perturbation theory.
It turns out, however, that for HERA energies the achieved accuracy is 
reasonably good; 
the estimation of the next term gives $\sim$ 30\% of the answer at not
too low $x$ (see the discussion in next section). 
Our final result for the DVCS amplitude with transversely polarized photons is
\footnote{In the leading logarithmic approximation it is not possible to
distinguish between $\alpha_s(Q)$ and $\alpha_s(\sqrt{|t|})$ -- to this end one 
needs to
use the NLO BFKL approximation\cite{nlobfkl} (see also \cite{cia}) 
which is beyond the scope 
of this paper.}
\begin{eqnarray}  
\lefteqn{ V^{AB}=}\label{fla42}\\
&&{2\over x}\left({\alpha_s(Q)\over\pi}\right)^2(\sum_{\rm flavors}e_q^2) 
F_1^{p+n}(t)\Bigg [
(e^A,e^B)_{\perp} v
+\left({2\over r_{\perp}^2}(e^A,r)_{\perp}(e^B,r)_{\perp}- 
(e^A,e^B)_{\perp}\right)v'
\Bigg ],
\nonumber
\end{eqnarray}
where
\begin{eqnarray}  
v(x,Q^2/t)&=&\left({1\over 2}\ln^2{Q^2\over |t|}+ 2\right)+{3\alpha_s(Q)\over \pi}
\ln{1\over x}\left({1\over 6} \ln^3{Q^2\over |t|} +2\ln{Q^2\over
|t|}-2+\zeta(3)\right)\nonumber\\ 
&~&+{1\over 2}\left({3\alpha_s(Q)\over \pi}\ln {1\over
x}\right)^2\left( {1\over 24} \ln^4{Q^2\over |t|}+\ln^2{Q^2\over |t|}
+2(\zeta(3)-1)\ln{Q^2\over |t|}+1.46\right)\label{fla43}\\
v'(x,Q^2/t)&=&
1+{3\alpha_s(Q)\over \pi}\ln {1\over x}+
{1\over 2}\left({3\alpha_s(Q)\over \pi}\ln {1\over x}\right)^2
\label{fla44}.
\end{eqnarray}
Note that the spin-dependent part $\sim v'$ does not contain any 
$\ln{Q^2\over |t|}$
 and is hence much smaller than the spin-independent part $\sim v$. For the
longitudinal polarization (\ref{fla13}) the amplitude is twist-suppressed as
$\simeq\sqrt {{|t|\over Q^2}}$ so we
have not calculated any terms beyond eq. (\ref{fla18}). In the numerical
analysis carried out in next sections we keep only the spin-independent
part of the amplitude
\begin{eqnarray}
V_{\perp}\equiv{1\over 4}\sum e_{\perp}^{A}e_{\perp}^{B}V^{AB}=
{2\over x}\left({\alpha_s(Q)\over\pi}\right)^2(\sum_{\rm flavors}e_q^2) 
F_1^{p+n}(t)v(x,Q^2,t)
\label{fla45}.
\end{eqnarray}

The above expressions give us the imaginary part of the DVCS amplitude. 
For the calculation of the 
DVCS cross section we need to know also the real
part of this amplitude
which can be estimated via the dispersion relation. 
For the positive-signature amplitude $H_{\perp}$ 
($\equiv{1\over 4}\sum e_{\perp}^{A}e_{\perp}^{B}H^{AB}$) we get\cite{bronzan} 
(see also \cite{strikfurt1})
\begin{equation}
{\rm Re} H_{\perp}(s)={\pi\over 2}\tan\left(s{d\over ds}\right) 
{\rm Im}H_{\perp}(s)
\label{fla46},
\end{equation}
which amounts to the substitution
\begin{equation}
\ln s \rightarrow {1\over 2}\big(\ln (-s-i\epsilon)+\ln s\big)
\label{fla47}
\end{equation}
in our amplitude (\ref{fla45}). Thus, the real part is
\begin{eqnarray}
\lefteqn{R\equiv {1\over \pi}{\rm Re} H_{\perp}={2\over x}\left({\alpha_s\over\pi}\right)^2
(\sum_{\rm flavors}e_q^2) 
(F_1^p(t)+F_1^n(t))r(x,Q^2,t)}\nonumber\\
&&r(x,Q^2,t)={\pi\over 2}\Bigg[{3\alpha_s\over\pi}\left({1\over 6} \ln^3{Q^2\over
|t|} +2\ln{Q^2\over |t|}-2+\zeta(3)\right)+
\nonumber\\
&&\left({3\alpha_s\over \pi}\right)^2\ln {1\over
x}\left( {1\over 24} \ln^4{Q^2\over |t|}+\ln^2{Q^2\over |t|}
+2(\zeta(3)-1)\ln{Q^2\over |t|}+1.46\right)\Bigg]
\label{fla48}.
\end{eqnarray}

\section{Comparison with the deep inelastic scattering}

It is instructive to compare the DVCS amplitude $V^{AB}$ given by 
eq. (\ref{fla3}) with 
the corresponding amplitude for the forward $\gamma^*$ scattering 
\begin{equation}
T^{AB}=ie^{A}_{\nu}e^{B}_{\mu}
\int dz e^{iqz}\langle p| T\{j^{\mu}(z)j^{\nu}(0)\} |p\rangle
\label{fla49}.
\end{equation}
The imaginary part of this amplitude is the total cross section for 
deep inelastic scattering (DIS)
\begin{eqnarray}
\lefteqn{{1\over \pi}{\rm Im} T^{AB}=W^{AB}=}\nonumber\\
&e^{A}_{\nu}e^{B}_{\mu}
\Bigg[\left({q_{\mu}q_{\nu}\over q^2}-g_{\mu\nu}\right)F_1(x,Q^2)+
{1\over pq}\left(p_{\mu}-q_{\mu}{pq\over q^2}\right)
\left(p_{\nu}-q_{\nu}{pq\over q^2}\right)F_2(x,Q^2)\Bigg]
\label{fla50}
\end{eqnarray}

For example  
$W^{AB}$ averaged over the transverse polarizations of the photons is
\begin{equation}
W_{\perp}\stackrel{\rm def}{\equiv} 
{1\over 4}\sum e_{\perp}^{A}e_{\perp}^{B}W^{AB}=F_1(x, Q^2)={1\over 2x}
F_2(x, Q^2)
\label{fla51}
\end{equation}
(at the leading twist level we have the Callan-Gross relation
$F_2=2xF_1$).
We will compare the imaginary part of the 
DVCS amplitude $V_{\perp}$ given by eq. (\ref{fla45}) 
to the result for $W_{\perp}$ calculated with the same accuracy. 
(We use the notation $W_{\perp}(x)$
rather than $F_1(x)$ in order to avoid confusion with $F_1(t)$).

  Similarly to the DVCS case, the DIS amplitude has the form 
(cf. eqs.(\ref{fla16},)(\ref{fla34}), and (\ref{fla39})):
\begin{eqnarray}
W_{\perp}&=&
{2g^2s\over \pi}\left(\sum e_q^2\right)
\int {d^2k_{\perp}\over 4\pi^2} {1\over k_{\perp}^4}
I_{\perp}(k_{\perp},0)\nonumber\\ 
&&\Bigg[1+
{3g^2\over 8\pi^3}\ln {1\over x}\int d^2k'_{\perp}  
 K(k_{\perp},k'_{\perp},0)
+\nonumber\\
&&{9g^4\over 128\pi^6}\ln^2 {1\over x}\int d^2k'_{\perp}\int d^2k"_{\perp}
K(k_{\perp},k"_{\perp},0){1\over (k''_{\perp})^2}
K(k"_{\perp},k'_{\perp},0)\Bigg]{1\over (k'_{\perp})^2}I_N(k'_{\perp},0)
\label{fla52}
\end{eqnarray}
where $I_{\perp}(k_{\perp},0)$ is the virtual photon impact factor 
averaged over
the transverse polarizations \cite{mes}
\begin{equation}
I_{\perp}(k_{\perp},0)=   
   {1\over 2}\int _{0}^{1} \frac {d\alpha}{2\pi }
   \int _{0}^{1} \frac {d\alpha'}{2\pi }
    {k_{\perp}^2(1-2\alpha\bar{\alpha})(1-2\alpha'\bar{\alpha'})\over 
      k_{\perp}^{2}\alpha'\bar{\alpha'}+Q^2\alpha'\alpha\bar{\alpha}}
   \label{fla53}
\end{equation}
The nucleon impact factor $I_N(k'_{\perp},0)$ cannot be calculated 
in perturbation theory since it is determined by the 
large-scale nucleon dynamics. However, we know the
asymptotics at large $k_{\perp}\gg m$
\begin{equation}
I_N(k_{\perp},0)\stackrel{k_{\perp}^2\gg m^2}{=}
F_1^{p+n}(0)=1
\label{fla54}
\end{equation}
Also, at $I_N(k_{\perp},0)\rightarrow 0$ at $k\rightarrow 0$ due to the 
gauge invariance. It seems reasonable to model this impact factor by 
the simple formula 
\begin{equation}
I_N(k_{\perp},0)={k_{\perp}^2\over k_{\perp}^2+m^2}
\label{fla55}
\end{equation}
which has the correct behavior both at large and small $k_{\perp}$.
With this model, the DIS amplitude (\ref{fla52}) takes the form
\begin{eqnarray}
\lefteqn{ W_{\perp}={F_2\over 2x}=}\label{fla56}\\
&&{4\over 3x}\left(\alpha_s(Q)\over \pi\right)^2(\sum_{\rm flavors}e_q^2) 
\Bigg[\left({1\over 2}\ln^2{Q^2\over m^2}+{7\over 6}\ln{Q^2\over m^2}+
{77\over 18}\right)+\nonumber\\
&&{3\alpha_s\over \pi}\ln {1\over x}\left({1\over 6}
 \ln^3{Q^2\over m^2}+{7\over 12}\ln^2{Q^2\over m^2}+
 {77\over 18}\ln{Q^2\over m^2}+{131\over 27}+2\zeta(3)\right)+{1\over 2}
 \left({3\alpha_s\over \pi}\ln {1\over x}\right)^2\nonumber\\
 &&
  \left({1\over 24}\ln^4{Q^2\over m^2}+{7\over 36}\ln^3{Q^2\over m^2}+
 {77\over 36}\ln^2{Q^2\over m^2 }+({131\over 27}+4\zeta(3))
 \ln{Q^2\over m^2}+
 {1396\over 81}-{\pi^4\over 15} +{14\over 3}\zeta(3)
  \right)
\Bigg ]\nonumber.
\end{eqnarray}
Note that the coefficients in front of leading logs of ${Q^2}$, determined by
the anomalous dimensions of twist-2 operators, coincide up to the factor
$2/3$. The graph of the model (\ref{fla56}) versus the
experimental data is presented in Fig. 7 for $Q^2=10$GeV$^2$
and  $Q^2=35$GeV$^2$ (we take
$\sum e_q^2={10\over 9}$).

\begin{figure}[htb]
\vspace{0cm}
\hspace{0cm}
\mbox{
\epsfxsize=7cm
\epsfysize=6cm
\hspace{0cm}
\epsffile{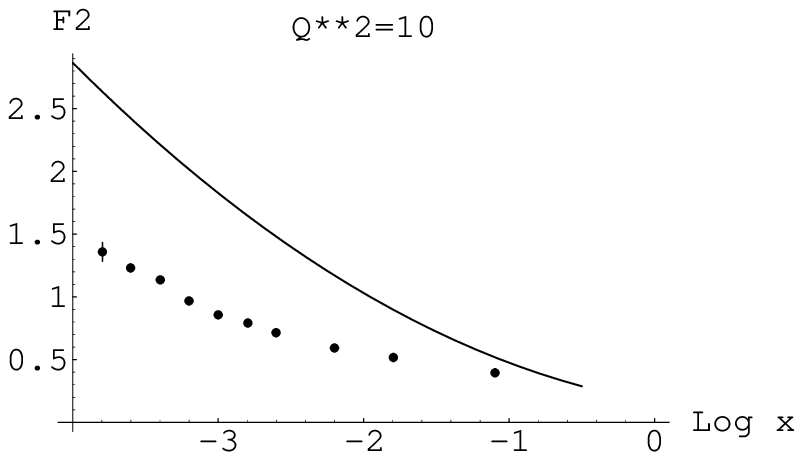}}
\hspace{1cm}
\mbox{
\epsfxsize=7cm
\epsfysize=6cm
\hspace{0cm}
\epsffile{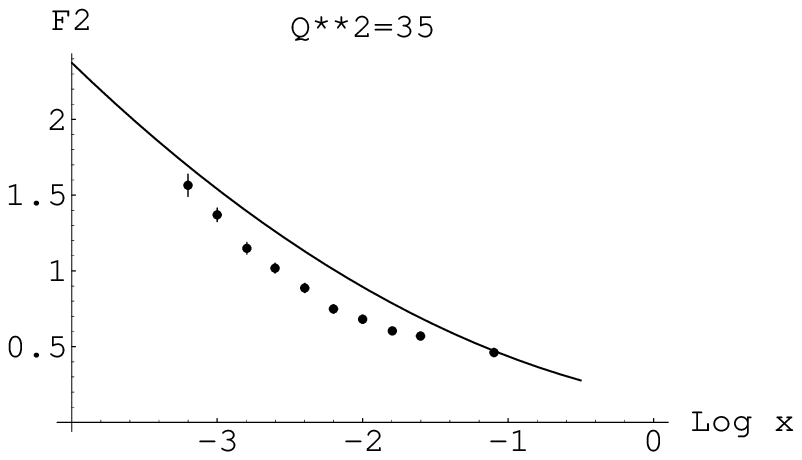}}
\vspace{0cm}
{\caption{\label{fig7} $F_2(x)$ from eq. (\ref{fla56}) versus experimental
data at $Q^2=10$GeV$^2$ and $Q^2=35$GeV$^2$ }} 
\end{figure}
\begin{figure}[htb]
\vspace{0cm}
\hspace{0cm}
\mbox{
\epsfxsize=7cm
\epsfysize=6cm
\hspace{0cm}
\epsffile{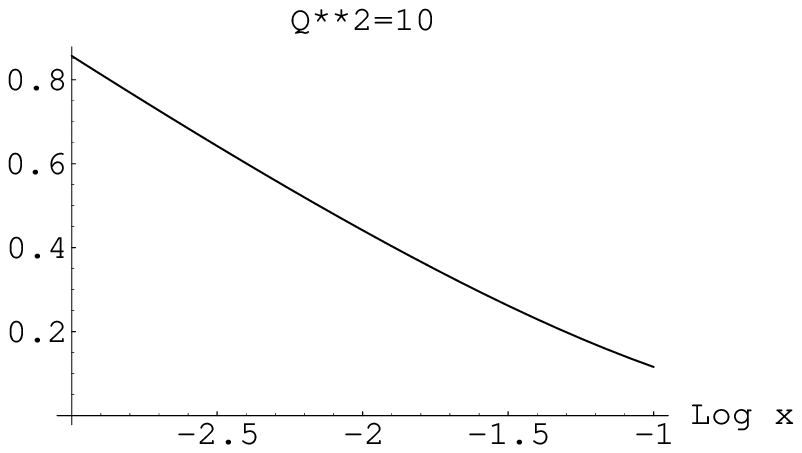}}
\hspace{1cm}
\mbox{
\epsfxsize=7cm
\epsfysize=6cm
\hspace{0cm}
\epsffile{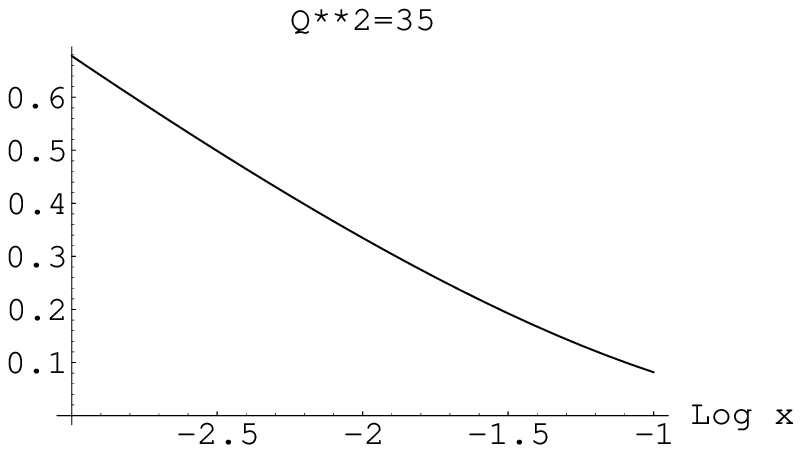}}
\vspace{0cm}
{\caption{\label{fig8} The ratio of the $\ln^3x$ term (\ref{fla57}) 
to eq. (\ref{fla56})
at $Q^2=10$GeV$^2$ and $Q^2=35$GeV$^2$}}  
\end{figure}

 
In the case of DIS it is possible to calculate explicitly the next term in
BFKL series (\ref{fla56})
\footnote{For DIS it is possible to write down the total BFKL sum as a 
Mellin integral 
and  unlike DVCS the integrals of impact factors with the BFKL 
eigenfunctions $(k_{\perp}^2)^{-{1\over 2}+i\nu}$ can be calculated explicitly.
Eqs. (\ref{fla56}) and (\ref{fla57}) correspond to the expansion of this 
explicit expression in powers of $\alpha_s\ln x$.}
. It has the form
\begin{eqnarray}
&&
{4\over 3x}\left(\alpha_s(Q)\over \pi\right)^2(\sum_{\rm flavors}e_q^2) 
\Bigg[
{1\over 6}\left({3\alpha_s\over \pi}\ln {1\over x}\right)^3
 \left({1\over 120}\ln^5{Q^2\over m^2}+
 {7\over 144}\ln^4{Q^2\over m^2}+{77\over 108}\ln^3{Q^2\over m^2}+
 ({131\over 54}+\right.\nonumber\\
 &&\left.+3\zeta(3))\ln^2{Q^2\over m^2 }+({1396\over 81}-
 {\pi^4\over 15}+7\zeta(3))
 \ln{Q^2\over m^2}+
 {4736\over 243}-{7\pi^4\over 90} +{77\over 3}\zeta(3)+6\zeta(5)
  \right)
\Bigg]\label{fla57}
\end{eqnarray}
The ratio of this 
$\left(\alpha_s\ln x\right)^3$ term to the sum of the first
three ones (\ref{fla56}) is presented in Fig. \ref{fig8} for $Q^2=10$GeV$^2$  
and $Q^2=35$GeV$^2$.
From
these graphs we see that the sum of the first tree terms gives the reliable
estimate of the DIS amplitude at not too low $x$ and it is  
expected that the same will also be true for DVCS amplitude 
\footnote{
At very small $x\sim 10^{-3}\div 10^{-5}$ the full BFKL result for 
$F_2$ in our model
 is growing more rapidly than Fig.
\ref{fig7}. 
On the other hand if one takes into account the NLO BFKL
corrections\cite{nlobfkl}\cite{cia} the result for $F_2$ at very small x
 goes well under the experimental points. This indicates 
 that at such  $x$ we need to unitarize 
 the BFKL pomeron, which is currently an unsolved problem. 
 (The best hope is to find the effective action for 
 the BFKL pomeron (see e.g. \cite{lipac},\cite{efek})).   
On the contrary, at ``intermediate'' $x\sim 0.1\div 0.001$, we 
see from Fig. \ref{fig7}
that, since
the corrections almost cancel each other, it makes sense to
take into account only a few first terms in BFKL series.
}.

\begin{figure}[htb]
\vspace{0cm}
\hspace{0cm}
\mbox{
\epsfxsize=6cm
\epsfysize=5cm
\hspace{0cm}
\epsffile{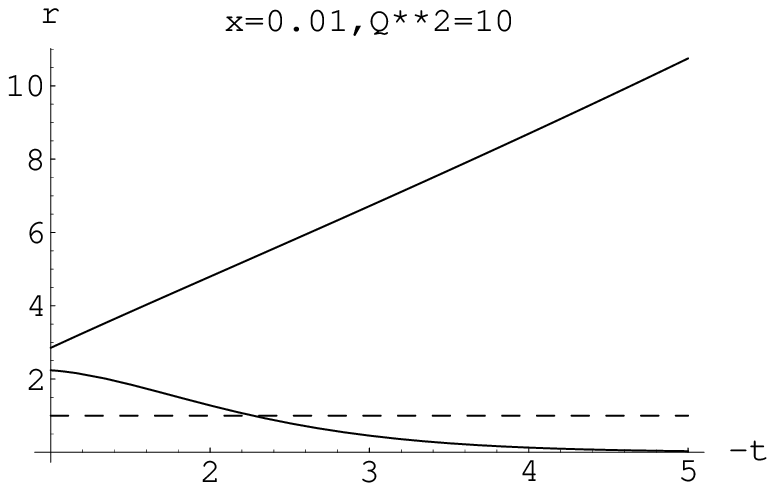}}
\hspace{2cm}
\mbox{
\epsfxsize=6cm
\epsfysize=5cm
\hspace{0cm}
\epsffile{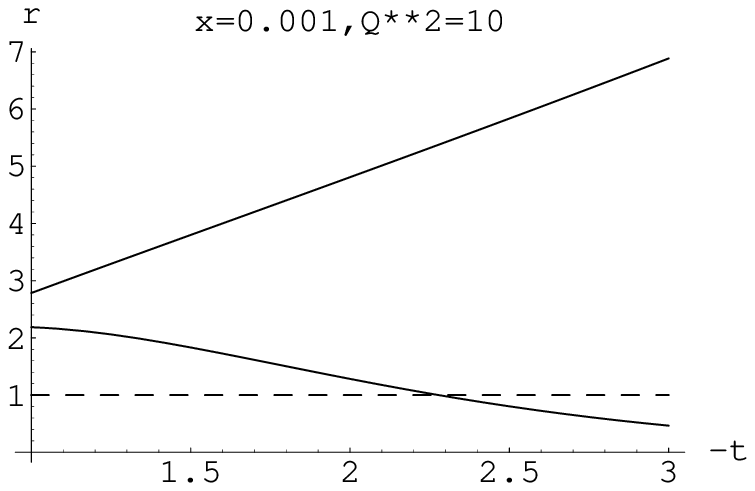}}\\

\vspace{0cm}
\mbox{
\epsfxsize=6cm
\epsfysize=5cm
\hspace{0cm}
\epsffile{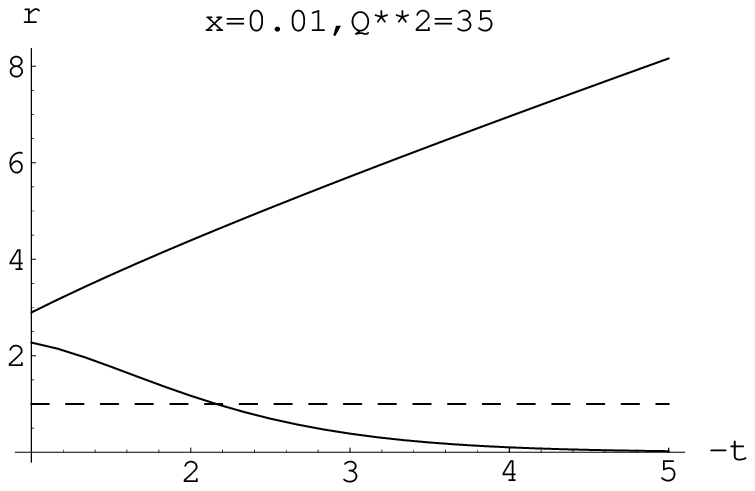}}
\hspace{2cm}
\mbox{
\epsfxsize=6cm
\epsfysize=5cm
\hspace{0cm}
\epsffile{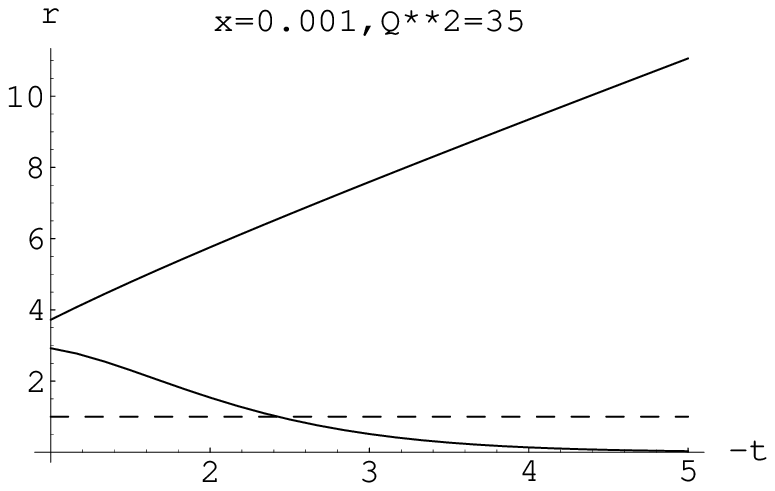}}
\vspace{0cm}
{\caption{\label{fig9} The ratio $V_1/V_{\perp}$ (lower curve) and
 $V_2/V_{\perp}$ (upper curve).}} \end{figure}

It is instructive to compare  the t-dependence of our DVCS amplitude
(\ref{fla43}) with the model used in the paper \cite{strikfurt1}
\begin{eqnarray}
V_1(x,t,Q^2)&=&{1\over R}F_1(x,Q^2)e^{bt/2}\label{fla58}\\
V_2(x,t,Q^2)&=&{1\over R}F_1(x,Q^2){1\over \big(1+{|t|\over 0.71}\big)^2}
\label{fla59}
\end{eqnarray}
where $R\simeq 0.5$ for our energies. (Literally, the model used in 
ref. \cite{strikfurt1} corresponds to $V_1$ but it is more natural to 
approximate the $t$ - dependence by the dipole formula\cite{private}). 
The comparison is shown in Fig. 
\ref{fig9} for $Q^2=10$GeV$^2$, $Q^2=35$GeV$^2$ and $x$=0.01, $x$=0.001.  
 
\section{DVCS cross section}

In order to estimate the cross section for DVCS at HERA kinematics
($Q^2>6$GeV$^2$ and $x<10^{-2}$)  we will use formulas from Ref.
\cite{strikfurt1} (see also Ref. \cite{dil}) with the trivial substitution 
${1\over 2x}F_2(x)R^{-1}e^{bt/2}\rightarrow V_{\perp}(x,Q^2,t)$. 
The expressions for
the DVCS cross section, the QED Compton (Bethe-Heitler) cross section, 
and the interference term have the form ($\bar{y}\equiv 1-y$)
\footnote
{The expression for the interference term from ref. \cite{strikfurt1} 
is corrected by factor 2 \cite{private}, \cite{belitsky}}:
 \begin{eqnarray}
{d\sigma^{\rm DVCS}\over dxdydtd\phi_r}&=& \pi\alpha^3x{1+\bar{y}^2\over Q^4y}
(V_{\perp}^2(x,Q^2,t)+R_{\perp}^2(x,Q^2,t))\label{fla60}\\
{d\sigma^{\rm QEDC}\over dxdydtd\phi_r}&=& {\alpha^3\over\pi x}
{y(1+\bar{y}^2)\over |t|Q^2\bar{y}} \left((F^p_1(t))^2+
{|t|\over 4m^2}(F_p^2(t))^2\right)\label{fla61}\\
{d\sigma^{\rm INT}\over dxdydtd\phi_r}&=& \mp 2\alpha^3
{(1+\bar{y}^2)\over Q^3\sqrt{\bar{y}|t|}} R_{\perp}(x,Q^2,t)
F^p_1(t)\cos\phi_r.
\label{fla62}
\end{eqnarray}
Here $y=1-{E'\over E}$ ($E$ and $E'$ are the incident and scattered
electron energies, respectively, as defined in the proton rest frame) 
and $\phi_r=\phi_e+\phi_N$ where $\phi_N$ is the azimuthal angle 
between the plane defined by $\gamma^*$ and the final state proton and the 
$x~-~z$ plane and $\phi_e$ is the azimuthal angle between the plane defined by
the initial and final state electron and $x~-~z$ plane 
(see Ref. \cite{strikfurt1}). As
mentioned above, we
approximate the Dirac and Pauli form factors of the proton by 
the dipole formulas (\ref{fla29}).
\begin{figure}[htb]
\vspace{0cm}
\hspace{0cm}
\mbox{
\epsfxsize=7cm
\epsfysize=6cm
\hspace{0cm}
\epsffile{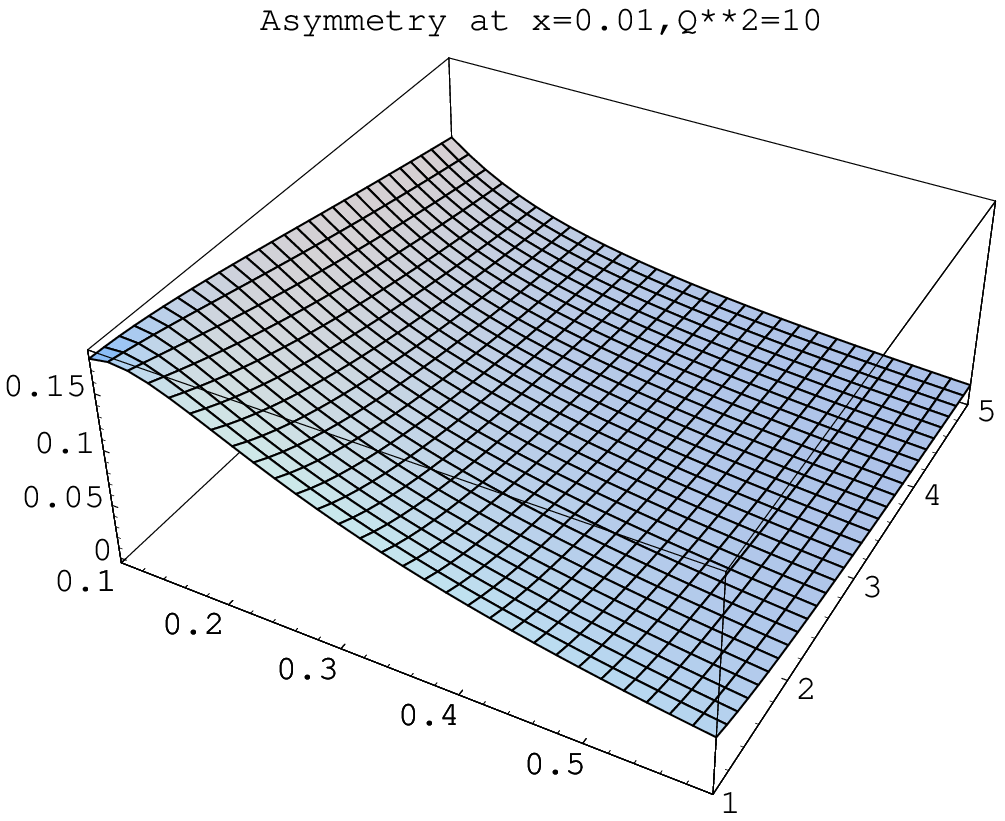}}
\hspace{1cm}
\mbox{
\epsfxsize=7cm
\epsfysize=6cm
\hspace{0cm}
\epsffile{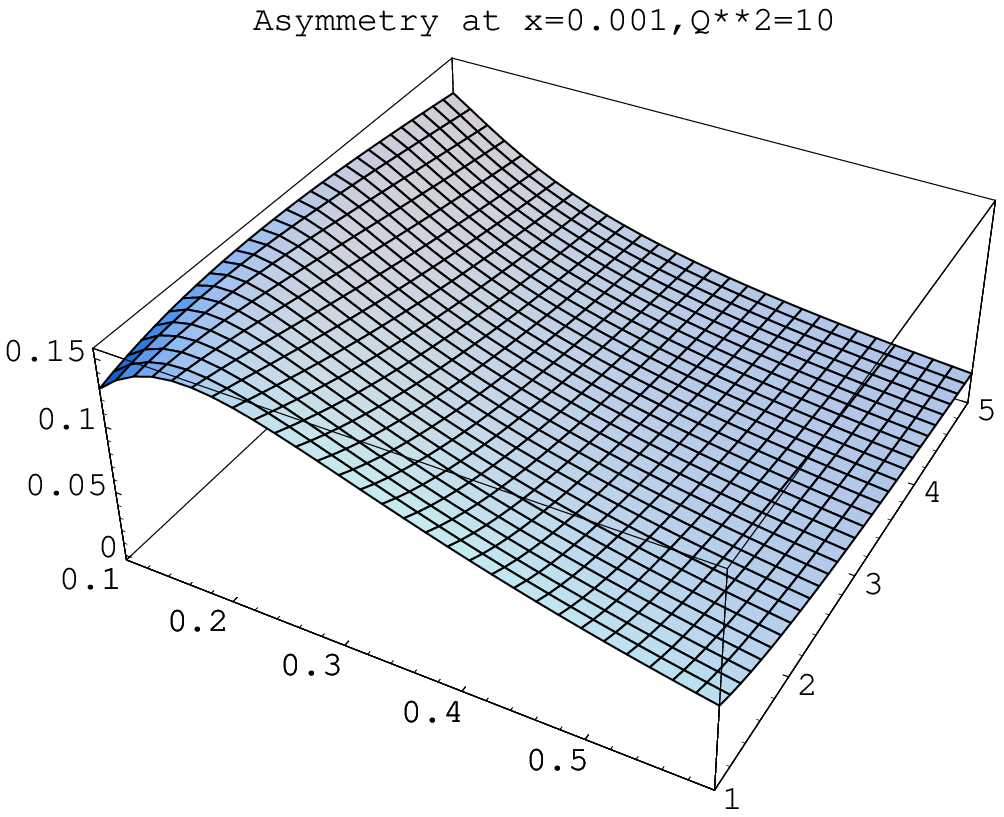}}
\vspace{1cm}
\mbox{
\epsfxsize=7cm
\epsfysize=6cm
\hspace{0cm}
\epsffile{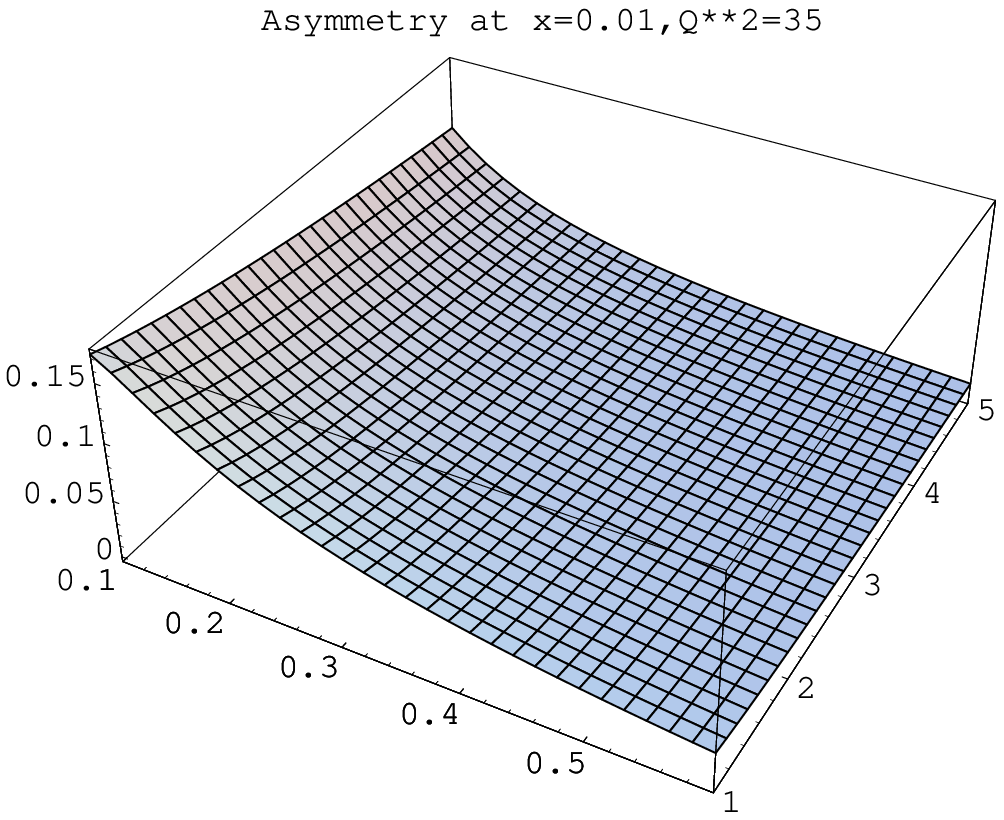}}
\hspace{1cm}
\mbox{
\epsfxsize=7cm
\epsfysize=6cm
\hspace{0cm}
\epsffile{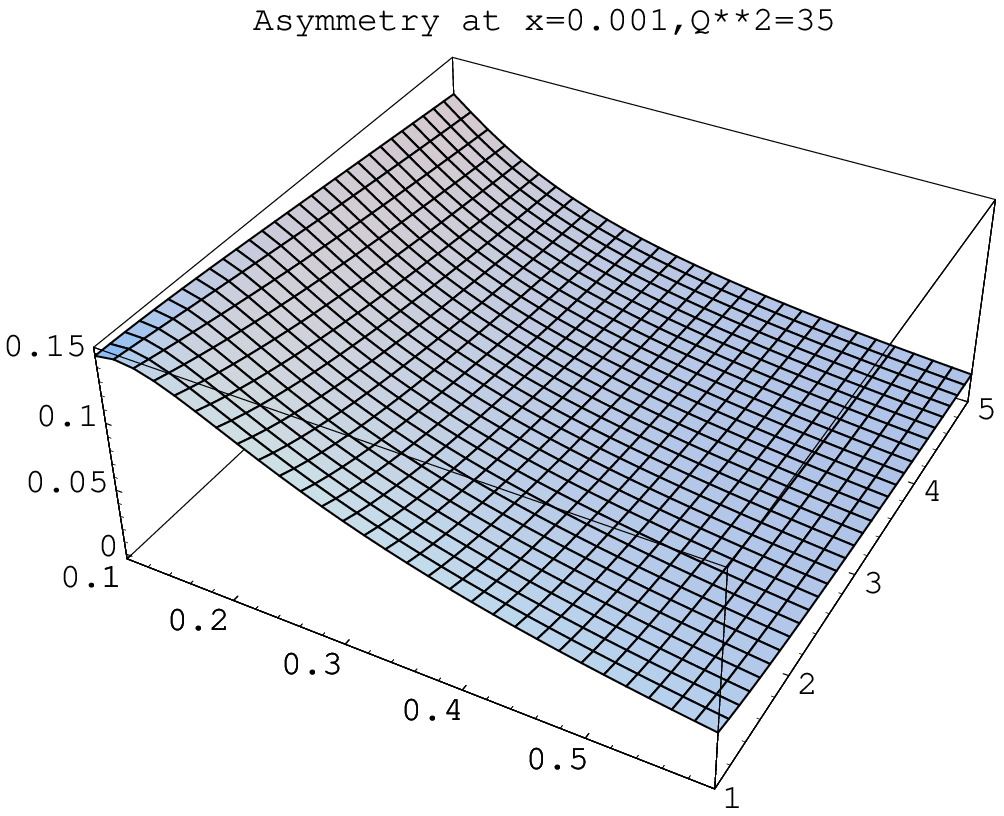}}
{\caption{\label{fig10} Asymmetry versus $y=0.1\div 0.6$ and $|t|=1\div 5$
GeV$^2$. }} 
\end{figure}

At first let us discuss the relative weight of the above cross sections.
We start with the asymmetry defined in ref. \cite{strikfurt2}
\begin{equation}
A={\int_{-\pi/2}^{\pi/2}d\phi_rd\sigma^{\rm DQI}-
\int_{\pi/2}^{3\pi/2}d\phi_rd\sigma^{\rm DQI}\over 
\int_{0}^{2\pi}d\phi_rd\sigma^{\rm DQI}}
\label{fla63}
\end{equation}
where 
\begin{equation} d\sigma^{\rm DQI}\equiv d\sigma^{\rm DVCS}+
d\sigma^{\rm QEDC}+d\sigma^{\rm INT}
\label{fla64}.
\end{equation} 
The asymmetry shows the relative importance of the interference term,
which is proportional to the real part of the DVCS amplitude.
In our
approximation the asymmetry is  
\begin{equation}
A(y,t)=
{4y\sqrt{{Q^2\over |t|\bar{y}}}(\sum e_q^2)\left({\alpha_s\over\pi}\right)^2
\left(1+2.8{|t|\over4m^2}\right)r\over
4\pi^2(\sum e_q^2)^2(v^2+r^2)\left({\alpha_s\over\pi}\right)^4
\left(1+{|t|\over 4m^2}\right)+
 {y^2Q^2\over \bar{y}|t|}\left(1+7.84{|t|\over 4m^2}\right)}
\label{fla65}
\end{equation}
The plots of asymmetry versus $y$ and $|t|$ are given by Fig. \ref{fig10}.

Second, we define the ratio of the DVCS and Bethe-Heitler cross
sections\cite{strikfurt1}
\begin{equation}
D(y,t)\equiv {d\sigma_{DVCS}\over d\sigma_{QEDC}}=
{4\pi^2(\sum e_q^2)^2(v^2+r^2)\left({\alpha_s\over\pi}\right)^4
\left(1+{|t|\over 4m^2}\right)\bar{y}{|t|\over Q^2}\over 
y^2\left(1+7.84{|t|\over 4m^2}\right)}
\label{fla66}
\end{equation}
This ratio is presented on Fig. 11. 
\begin{figure}[htb]
\vspace{0cm}
\hspace{0cm}
\mbox{
\epsfxsize=7cm
\epsfysize=6cm
\hspace{0cm}
\epsffile{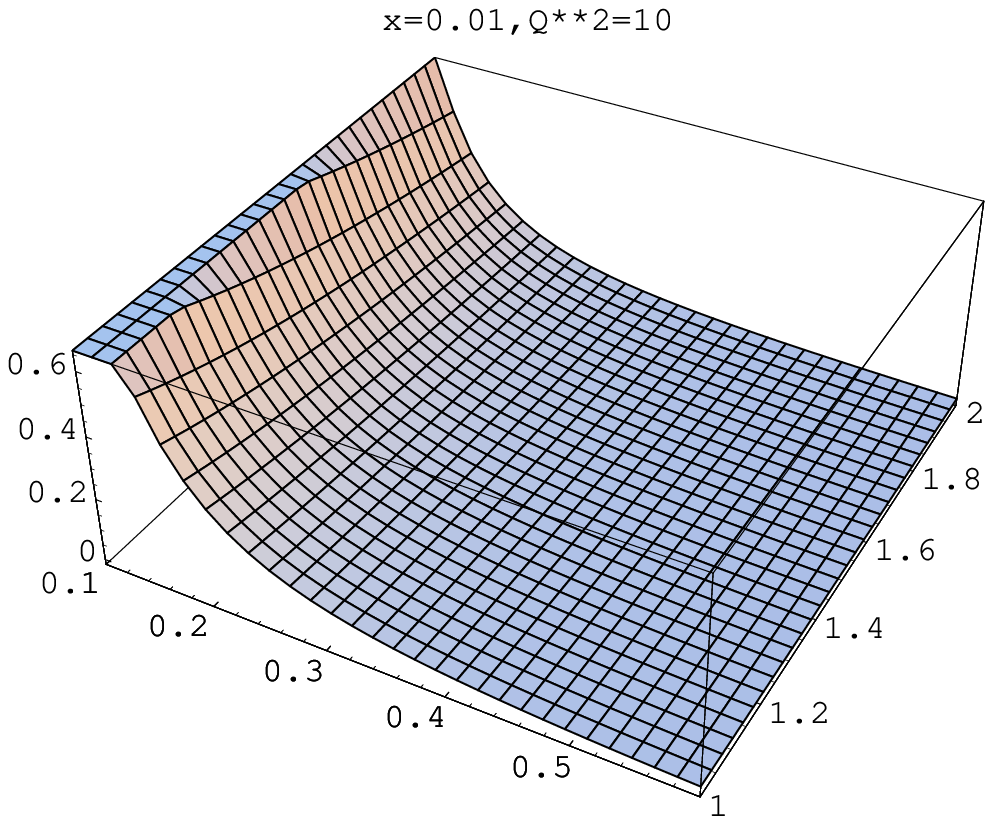}}
\hspace{1cm}
\mbox{
\epsfxsize=7cm
\epsfysize=6cm
\hspace{0cm}
\epsffile{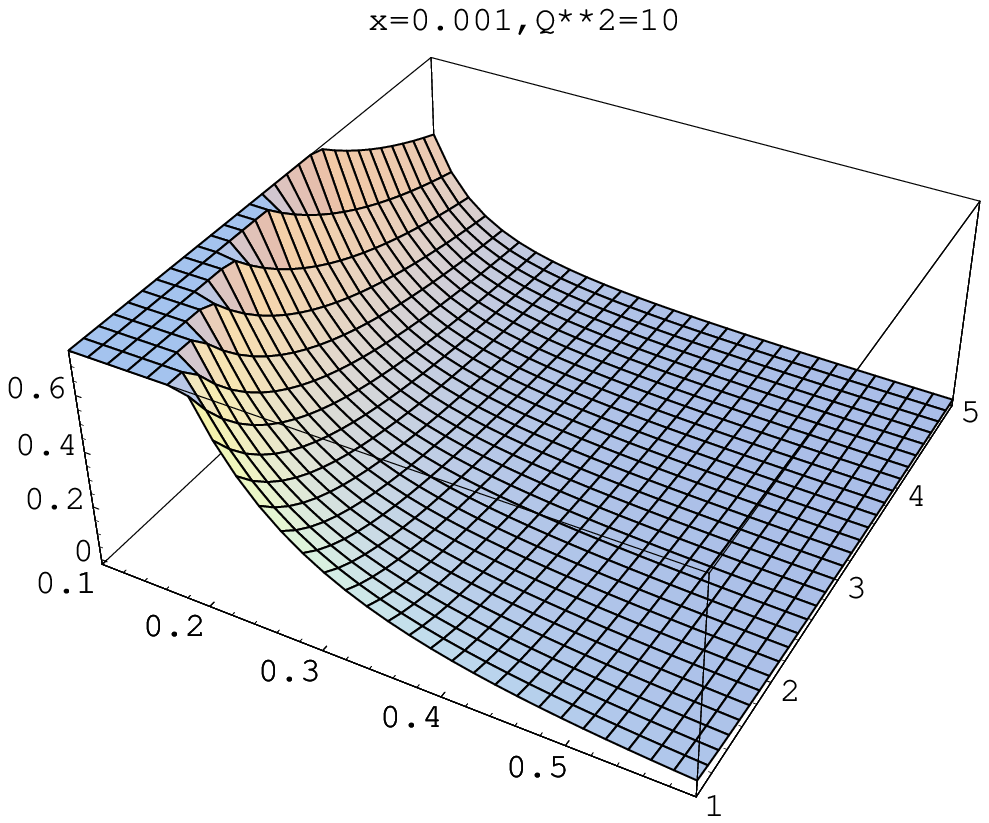}}
\vspace{1cm}
\mbox{
\epsfxsize=7cm
\epsfysize=6cm
\hspace{0cm}
\epsffile{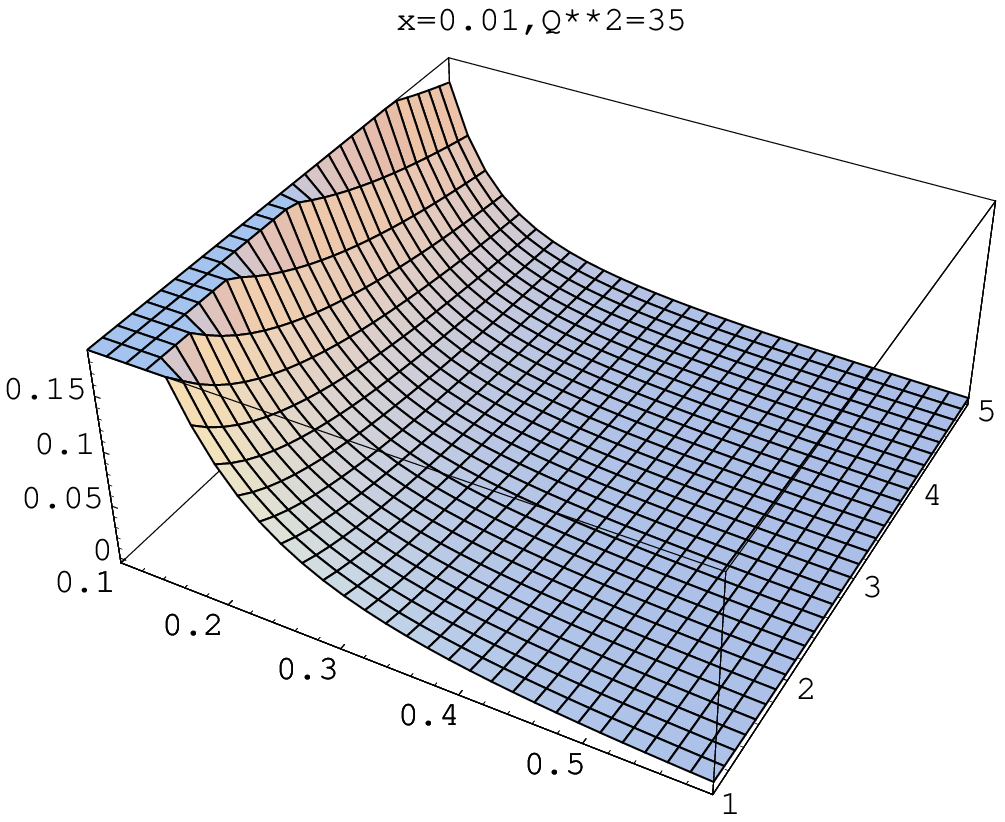}}
\hspace{1cm}
\mbox{
\epsfxsize=7cm
\epsfysize=6cm
\hspace{0cm}
\epsffile{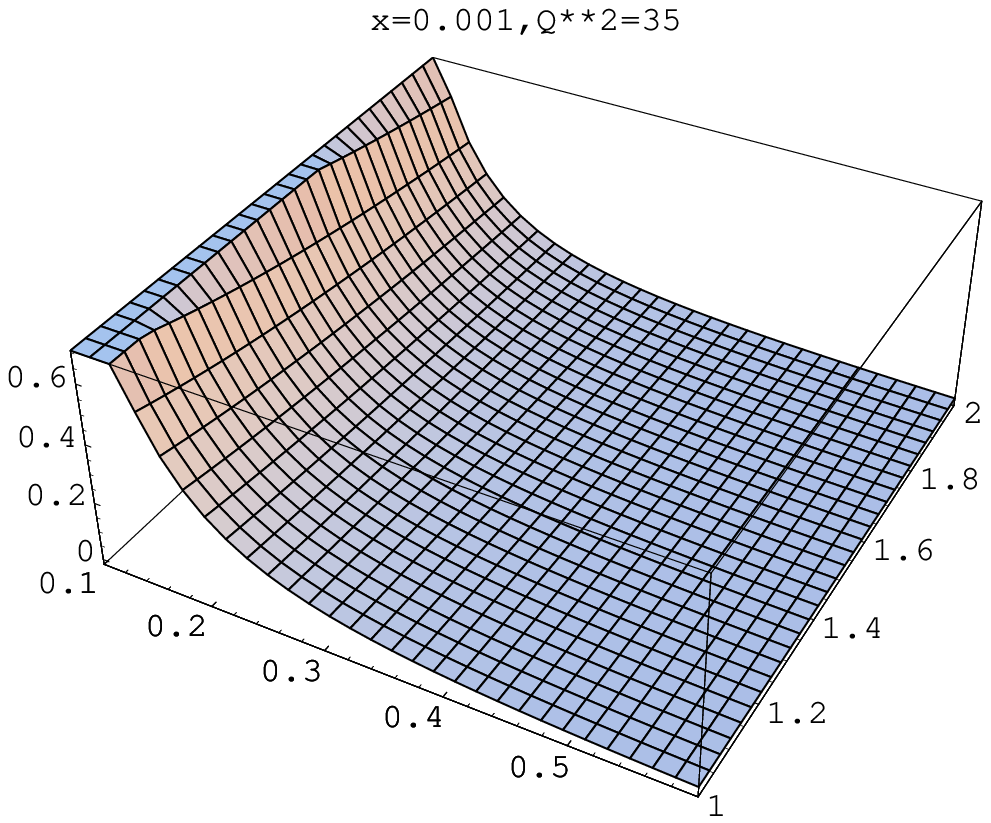}}
{\caption{\label{fig11} The ratio $D(x,Q^2/t)$ versus $y=0.1\div 0.6$ 
and $|t|=1\div 5$
GeV$^2$}} 
\end{figure} 
We see that there is a sharp dependence on $y$; 
at $y>0.2$ the DVCS part is negligible in comparison
to Bethe-Heitler background whereas at $y<0.05$ 
the QEDC background is small in comparison to DVCS.

 Finally let us estimate the relative weight of the DVCS signal (starting
 from $|t|=1$ GeV$^2$) as compared to the DIS
background. We define (cf. ref. \cite{strikfurt1}) 
\begin{eqnarray}
\lefteqn{R_{\gamma}=
{\sigma(\gamma^*+p\rightarrow \gamma+p)\over \sigma(\gamma^*+p\rightarrow
\gamma^*+p)}\simeq}\nonumber\\
&{4\pi\alpha\over Q^2F_2(x,Q^2)}\left({\alpha_s\over \pi}\right)^4
\left(\sum e_q^2\right)^2\int^{Q^2}_1 dt
\left(F_1^{p+n}(t)\right)^2(v^2(x,Q^2/t)+r^2(x,Q^2/t))
\label{fla67}
\end{eqnarray}
At $Q^2=10$GeV$^2$ we find $R_\gamma=1.56\times 10^{-5}$ for $x=0.01$ and
$R_\gamma=2.36\times 10^{-5}$ for $x=0.001$, while for $Q^2=35$GeV$^2$ 
 we find $R_\gamma=0.62\times 10^{-5}$ for $x=0.01$ and
$R_\gamma=0.71\times 10^{-5}$ for $x=0.001$.
\begin{figure}[htb]
\vspace{0cm}
\hspace{0cm}
\mbox{
\epsfxsize=7cm
\epsfysize=6cm
\hspace{0cm}
\epsffile{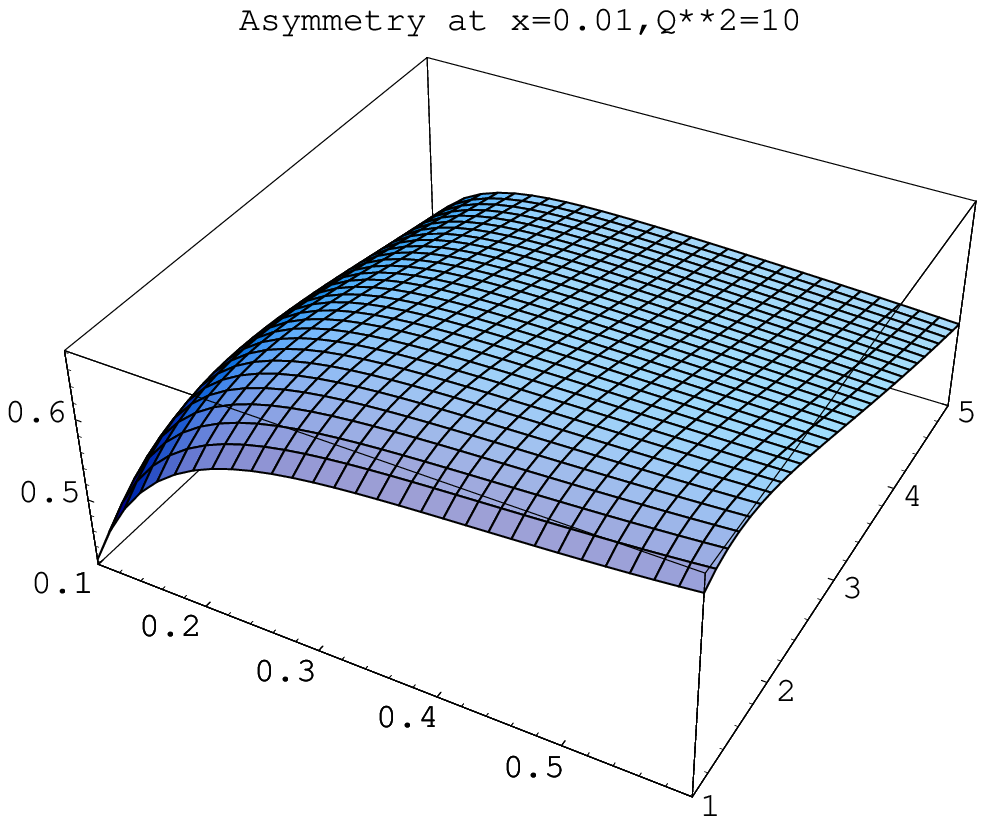}}
\hspace{1cm}
\mbox{
\epsfxsize=7cm
\epsfysize=6cm
\hspace{0cm}
\epsffile{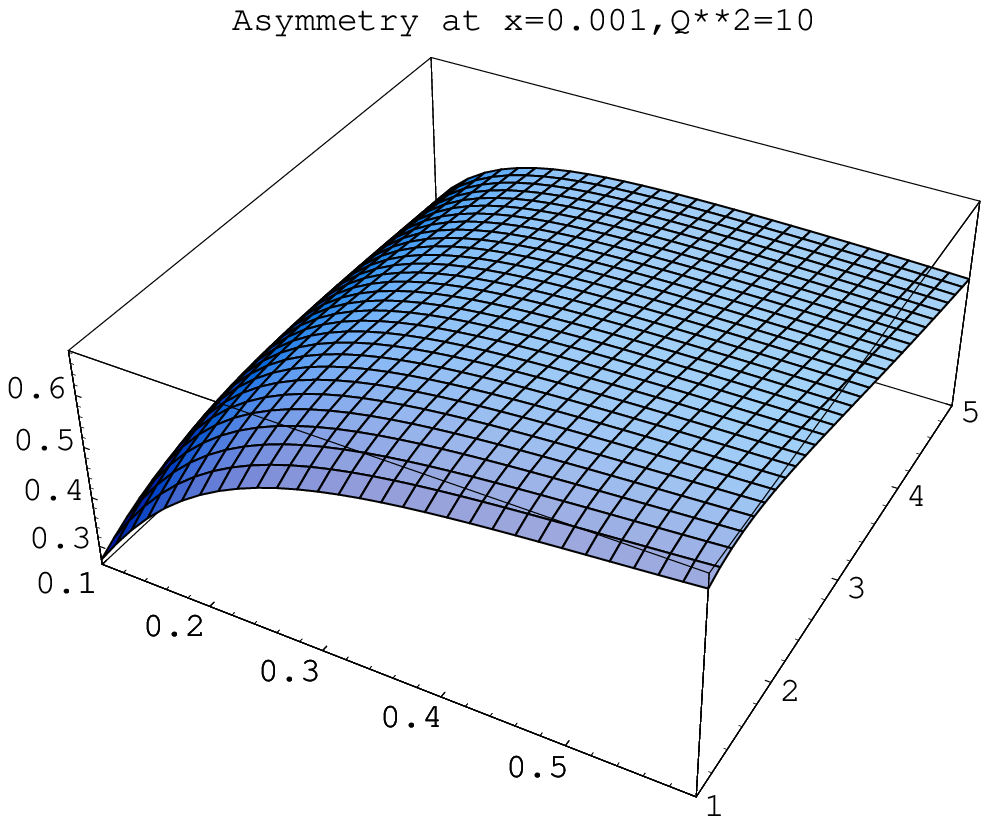}}
\vspace{1cm}
\mbox{
\epsfxsize=7cm
\epsfysize=6cm
\hspace{0cm}
\epsffile{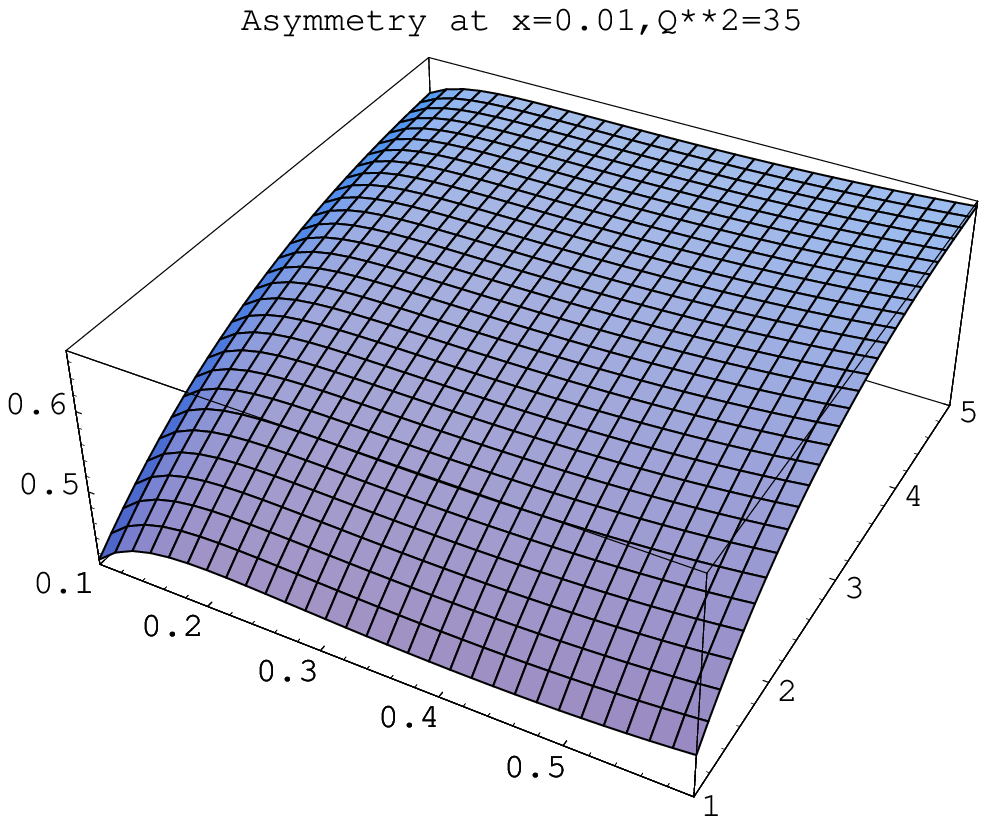}}
\hspace{1cm}
\mbox{
\epsfxsize=7cm
\epsfysize=6cm
\hspace{0cm}
\epsffile{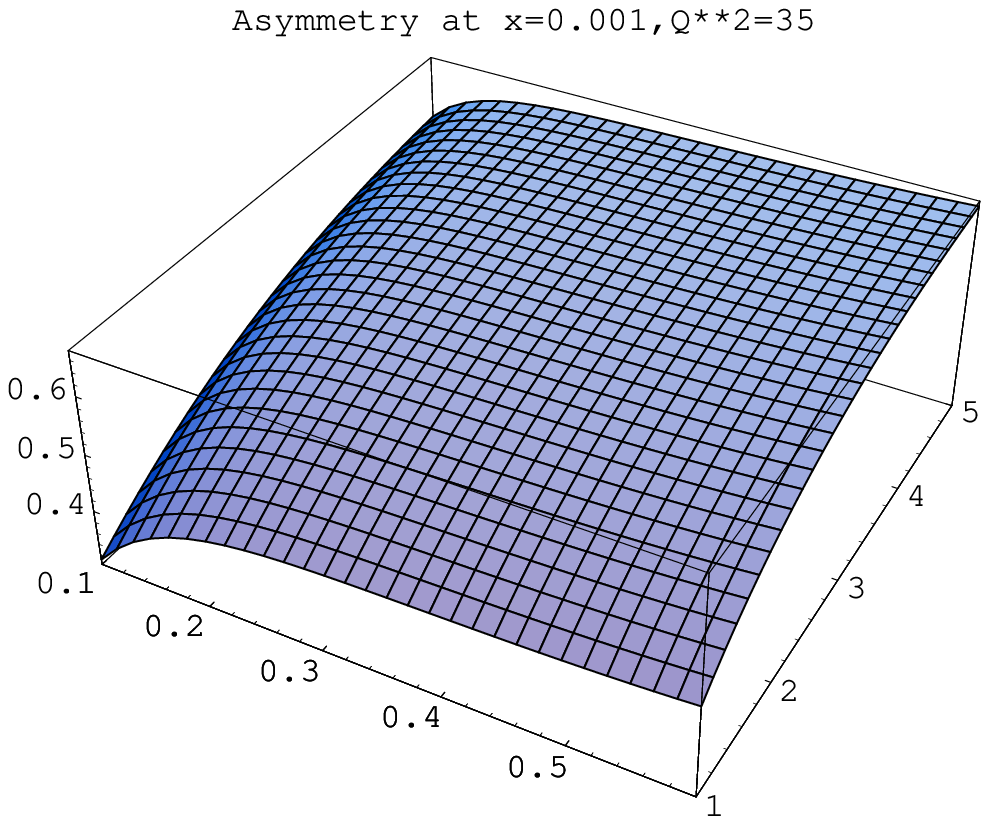}}
{\caption{\label{fig12} Asymmetry with the correction factor (68). }} 
\end{figure}

The expressions (60)-(62) are correct if $Q^2\ll |t|$ up to
$O({|t|\over Q^2})$ accuracy with the notable exception of 
the correction $O({\sqrt{|t|}\over Q})$ coming from the expansion of
electron propagator in the u-channel of the Bethe-Heitler 
amplitude. As suggested in ref.  \cite{belitsky}, at intermediate $t$ 
one can keep the propagator in unexpanded form 
(and expand the rest of the amplitude, as we have done above). 
This amounts to the replacement
\begin{equation}
\bar{y}\rightarrow  
\bar{y}\Bigg[(1+{|t|\over Q^2\bar{y}})(1+{|t|\bar{y}\over Q^2})-
2{(2-y)\over\sqrt{\bar{y}}}\sqrt{{|t|\over Q^2}}\cos\phi_r+
4{|t|\over Q^2}\cos^2\phi_r\Bigg]
\end{equation}
in the numerator in eqs. (61) and (62) (see ref. \cite{dil}). 
\begin{figure}[htb]
\vspace{0cm}
\hspace{0cm}
\mbox{
\epsfxsize=7cm
\epsfysize=6cm
\hspace{0cm}
\epsffile{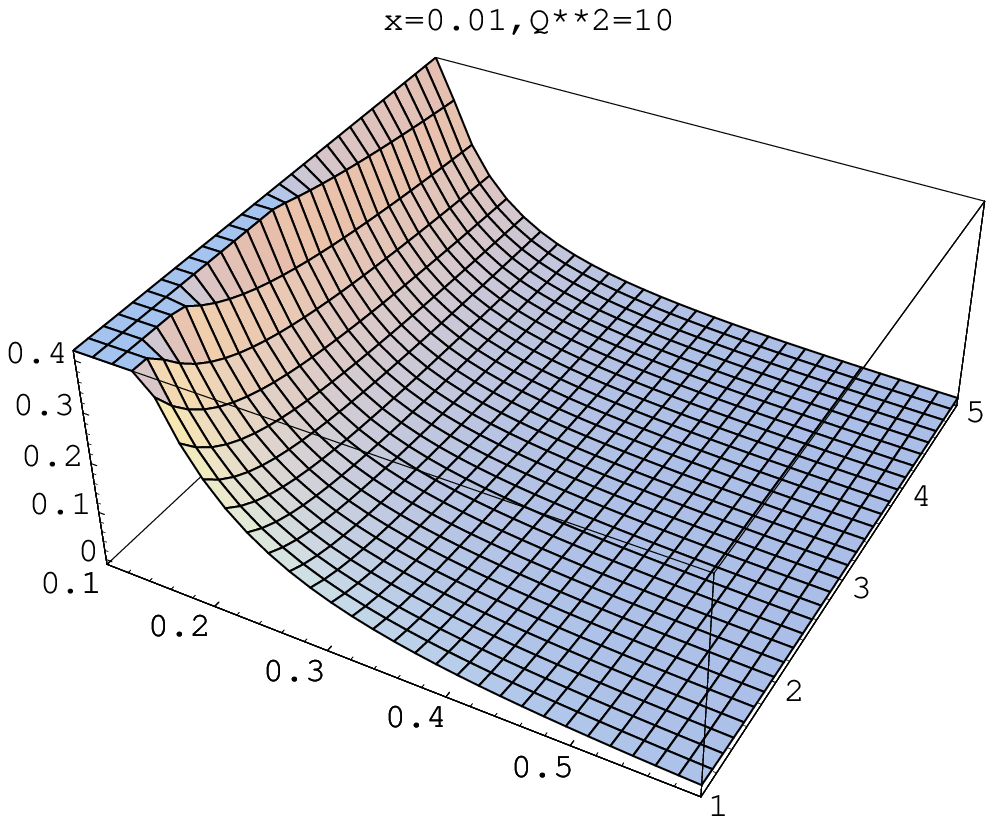}}
\hspace{1cm}
\mbox{
\epsfxsize=7cm
\epsfysize=6cm
\hspace{0cm}
\epsffile{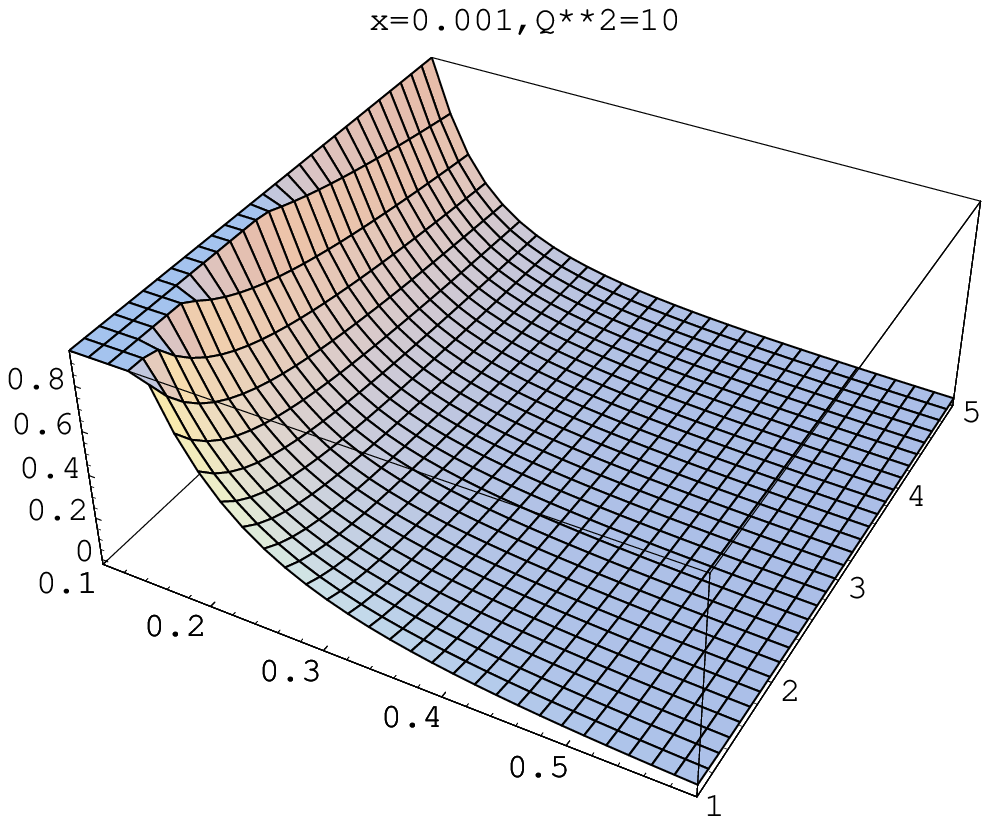}}
\vspace{1cm}
\mbox{
\epsfxsize=7cm
\epsfysize=6cm
\hspace{0cm}
\epsffile{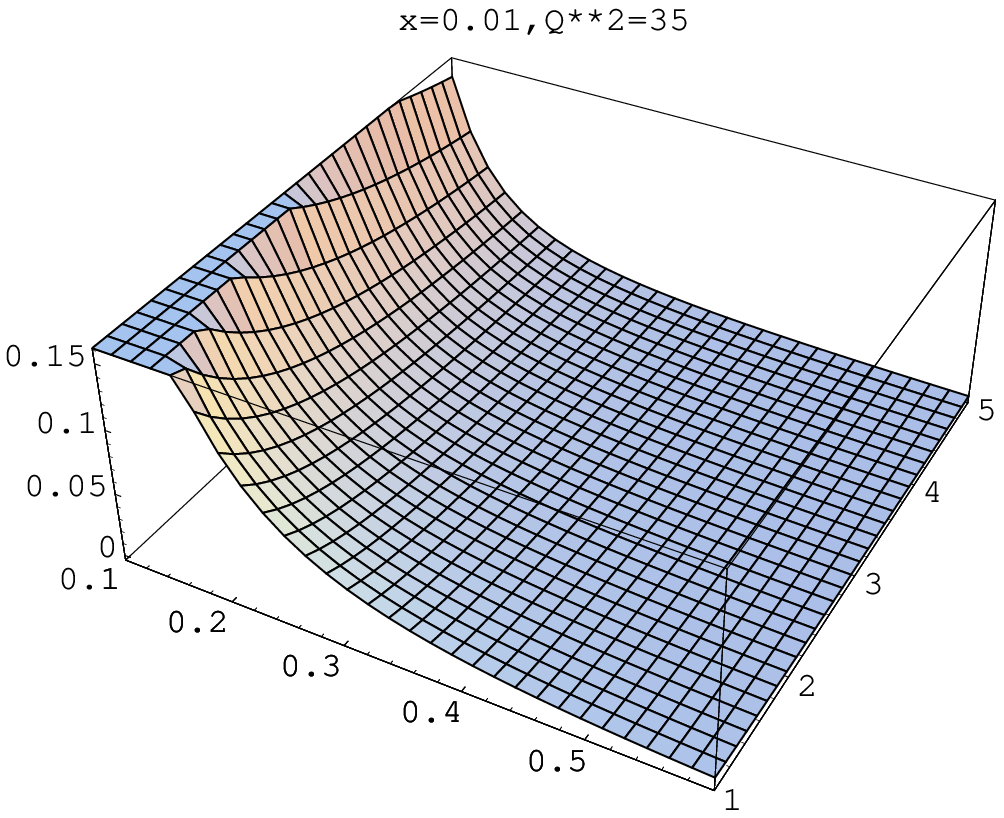}}
\hspace{1cm}
\mbox{
\epsfxsize=7cm
\epsfysize=6cm
\hspace{0cm}
\epsffile{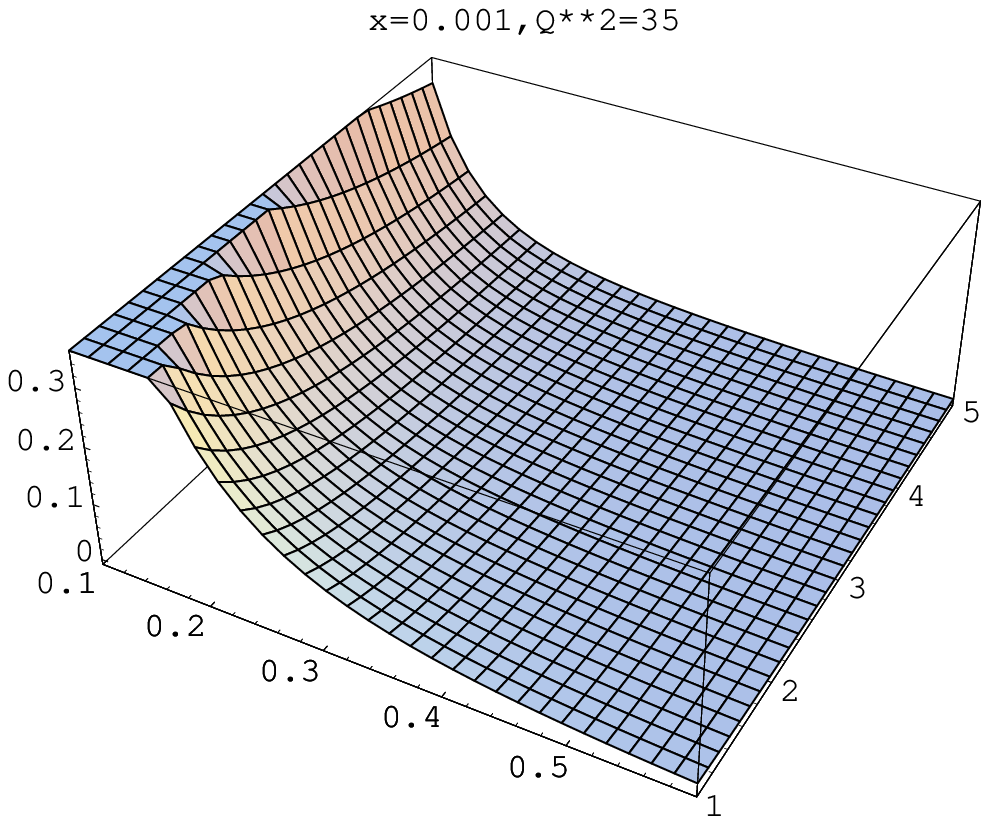}}
{\caption{\label{fig12} The ratio $D(x,Q^2/t)$ with the correction factor
(68). }}  \end{figure}
The resulting asymmetry (63) is presented in Fig. 12. 
We see that the correction factor (68) crucially changes  
the behavior of the asymmetry due to the fact that it restores the azimuthal
dependence of the QEDC amplitude which was not taken into account in eqs.
(60-62). In order to find asymmerty at these $Q^2$ 
and $t$ with greater accuracy one should take into
account other twist-4 contributions as well.  On the contrary, the ratio
$D(x,Q^2/t)$ does not change much (see Fig. 13) so we hope that our 
leading-twist results for the ratio presented in Fig. 11 are reliable. 

\section{Conclusion}
 The DVCS in the kinematical region (\ref{fla1}) is probably the best place 
 to test the momentum transfer dependence of the BFKL pomeron.
 Without this dependence, the  model
 (\ref{fla59}) would be exact, hence the upper curves in Fig. \ref{fig9} 
 indicate how important is the $t$-dynamics of the pomeron. We see 
 that the $t$-dependence of the BFKL pomeron changes the cross section
  at $t>2$GeV$^2$ by orders of magnitude and therefore it should be be 
  possible to 
 detect it.
 
 The pQCD calculation of the DVCS amplitude in the region (\ref{fla1}) is
 in a sense  
 even more reliable than the calculation of usual DIS amplitudes since
 it does not rely on the models for nucleon parton distributions . Indeed,
 all the non-perturbative
 nucleon input is contained in the Dirac form factor of the nucleon
 \footnote{
 There are, of course, the non-perturbative corrections to the 
 BFKL pomeron itself.  
 At present, it is not clear how to take them into account.}, 
  which is known to a pretty good accuracy. (Of course
 any reasonable models of nucleon parton distributions such as (\ref{fla24})
 should reproduce the form factor after integration over $X$).
 
 Finally, let us discuss uncertainties in our approximation and
 possible ways to improve it. One obvious improvement would be to
 calculate (at least numerically) the next 
 $\sim\left(\alpha_s\ln x\right)^3$ term in the BFKL series for 
 the DVCS amplitude. Hopefully, it will be as small as the corresponding 
 calculation of the DIS amplitude suggests. 
 Second, there are non-perturbative corrections to the BFKL pomeron
 which we mention above. These non-perturbative corrections
 correspond to the situation like the ``aligned jet model'' when one
 of the two gluons in Fig. (\ref{fig1}) is soft and all the momentum
 transfers through the other gluon. It is not 
 clear how to take these corrections into account, but one should expect
 them to be smaller than the corresponding corrections to $F_2(x)$ which
 come from two  non-perturbative gluons in Fig. 1 (in other words, from 
 the ``soft pomeron'' 
 contribution to $F_2(x)$).

 The biggest uncertainty in our calculation is the argument
 of coupling constant $\alpha_s$ which we take to be $Q^2$. As we
 mention above, it is not possible to fix the argument of 
 $\alpha_s$ in the LLA, so we could have used $\alpha_s(|t|)$ instead. 
 We hope to overcome
 this difficulty by using the results of NLO BFKL in our future work.

\vspace{-0.5cm}
\section*{Acknowledgments}

The authors are grateful 
to A.V. Belitsky, C.E. Hyde-Wright, I.V. Musatov,  A.V. Radyushkin, and M.I.
Strikman  for valuable discussions.
 This work was supported by DOE contract DE-AC05-84ER40150 under which the
Southeastern Universities Research Association (SURA) operates
the Thomas Jefferson National Accelerator Facility.


\section*{References}
\begin{thebibliography}{99}

\bibitem{leipzig}
D.Muller, D.Robaschik, B. Geyer, F.M. Dittes, and  J. Horejsi,
\Journal{\em Fortschr.Phys.}{42}{101}{1994} 


\bibitem{ji}
 X. Ji, \Journal{\PLB}{78}{610}{1997}, \Journal{\PRD}{55}{7114}{1997}

\bibitem{rad1}
A.V. Radyushkin,\Journal{\PLB}{380}{417}{1996},\Journal{\PLB}{385}{333}{1996}

\bibitem{bartels}
 J. Bartels and M. Loewe, 
 \Journal{\em Z. Phys. C}{12}{263}{1982} 

\bibitem{bfkl}
  V.S. Fadin, E.A. Kuraev, and L.N. Lipatov, \Journal{\PLB}{60}{50}{1975};\\
 I.I. Balitsky and L.N. Lipatov, \Journal{\em Sov. Journ. Nucl. Phys.} 
{28}{822}{1978}  

\bibitem{zeus}
  [ZEUS Collaboration], {\it Observation of DVCS in 
  $e^+$ Interactions at HERA}, paper submitted to EPS HEP99 
  Conference, Tampere, 1999 (see also http://www-zeus.desy.de)
  
\bibitem{strikfurt1}
  L.L. Frankfurt, A. Freund, and M.I. Strikman,
\Journal{\PRD}{58}{114001}{1998}, Erratum-
\Journal{\PRD}{59}{119901}{1999}

\bibitem{guvan} P.A.M. Guichon, M. Vanderhaeghen,  
\Journal{Prog.Part.Nucl.Phys.}{41}{125}{1998} 

\bibitem{rad2}
A.V. Radyushkin, \Journal{\PRD}{56}{5524}{1997}

\bibitem{collins}
  J.C. Collins, L.L. Frankfurt, and M.I. Strikman,
\Journal{\PRD}{56}{2982}{1997}


\bibitem{lobzor}
  L.N. Lipatov, 
\Journal{\em Phys. Reports} {286}{131}{1997}.

\bibitem{ifak}
H. Cheng and T.T. Wu, \Journal{\em Phys. Rev.}{182}{1852}{1969},
\Journal{\PRD}{1}{2775}{1970}; V.N. Gribov et al., \Journal{\em Sov. 
J. Nucl. Phys.}{12}{543}{1971}.

\bibitem{ing}
  I. Balitsky, 
  \Journal{\NPB}{463}{99}{1996}.

\bibitem{bal83}
  I. Balitsky, \Journal{\PLB}{124}{230}{1983}.

\bibitem{prof} 
A.V. Radyushkin,
\Journal{\PRD}{58}{114008}{1998}

\bibitem{nlobfkl}
 L.N. Lipatov ans V.S. Fadin, 
\Journal{\PLB}{429}{127}{1998}, \Journal{\NPB}{477}{767}{1997},
\Journal{\NPB}{406}{259}{1993}

\bibitem{cia}
 G. Carnici and M. Ciafaloni,
\Journal{\PLB}{412}{396}{1997}, \Journal{\PLB}{430}{349}{1998}

\bibitem{bronzan}
  J.B. Bronzan, G.L. Kane, and U.P. Sukhatme, \Journal{\PLB}{49}{272}{1974}.

\bibitem{mes}
  I.I. Balitsky and L.N. Lipatov, 
\Journal{\em JETP Letters} {30}{355}{1979}.

\bibitem{lipac} 
L.N. Lipatov,
\Journal{\NPB}{452}{369}{1996}

\bibitem{efek}
  I. Balitsky, \Journal{\PRD}{014020}{60}{1999}.

\bibitem{private}
 M.I. Strikman, private communication.

\bibitem{dil}
M. Diehl, T. Gousset, B. Pire , J.P. Ralston 
\Journal{PLB}{411}{193}{1997}.

\bibitem{strikfurt2}
   L.L. Frankfurt, A. Freund, and M.I. Strikman,
\Journal{\PLB}{460}{417}{1999}

\bibitem{belitsky}
 A.V. Belitsky, D. Mueller, L. Niedermeier, and A. Shaefer, hep-ph/0004059.


\end{thebibliography}
\end{document}